\documentclass[a4paper,11pt]{article}
\usepackage{mathrsfs,amstext,amsmath,amssymb,amsfonts,cite,xcolor,epsf,epsfig,epstopdf,graphics,graphicx,mathtools,physics,verbatim,setspace,subfig,soul}
\usepackage{tcolorbox}
\def\thefootnote{*\arabic{footnote}}
\definecolor{ultramarine}{rgb}{0.07, 0.04, 0.56}
\definecolor{DarkBlue}{rgb}{0.0,0.0,0.4}
\definecolor{cadmiumgreen}{rgb}{0.0, 0.42, 0.24}
\definecolor{indigo(dye)}{rgb}{0.0, 0.25, 0.42}
 
\usepackage[linktocpage=true,breaklinks]{hyperref}
\usepackage[normalem]{ulem}
\hypersetup{
	colorlinks=true,
	citecolor=ultramarine,
	linkcolor=DarkBlue,
	urlcolor=indigo(dye),
}

\numberwithin{equation}{section}
\graphicspath{{figures/}}

\newcommand{\nc}{\newcommand}
\nc{\ba}{\begin{eqnarray}}
\nc{\ea}{\end{eqnarray}}

\newcommand{\p}{{\partial}}
\newcommand{\Mpl}{M_{\rm Pl}}

\newcommand{\eq}[1]{\begin{equation}#1\end{equation}}
\newcommand{\eqa}[1]{\begin{align}#1\end{align}}
\newcommand{\spl}[1]{\begin{split} #1 \end{split}}
\newcommand{\fg}[1]{\begin{figure}[tbp]\centering #1 \end{figure}}
\newcommand{\bxeq}[1]{\begin{tcolorbox}[colback=gray!12,colframe=black!40,boxrule=0.15mm]\begin{equation}#1\end{equation}\end{tcolorbox}}

\newcommand{\vp}{\varphi}

\newcommand{\mhn}[1]{{\color{blue} MHN: #1}}

\definecolor{livingcoral}{RGB}{74,144,226}
\definecolor{redd}{RGB}{65,117,5}

\usepackage[framemethod=TikZ]{mdframed}
\mdfsetup{roundcorner=3pt}   
\newmdenv[
skipabove=5pt,
skipbelow=7pt,
rightline=false,
leftline=false,
topline=false,
bottomline=false,
backgroundcolor=gray!15,
innerleftmargin=5pt,
innerrightmargin=5pt,
innertopmargin=7pt,
innerbottommargin=7pt,
leftmargin=0cm,
rightmargin=0cm,
linewidth=4pt]{eBox}

\newmdenv[
skipabove=5pt,
skipbelow=7pt,
rightline=false,
leftline=false,
topline=false,
bottomline=false,
backgroundcolor=olive!20,
innerleftmargin=5pt,
innerrightmargin=5pt,
innertopmargin=7pt,
innerbottommargin=7pt,
leftmargin=0cm,
rightmargin=0cm,
linewidth=4pt]{eBox2}

\makeatletter
\def\blfootnote{\xdef\@thefnmark{}\@footnotetext}
\makeatother

\begin{document}

\begin{titlepage}
\setcounter{page}{1} \baselineskip=15.5pt \thispagestyle{empty}
\begin{flushright} {\footnotesize MIT-CTP/5293}  \end{flushright}
\vspace{5mm}
\vspace{0.5cm}
\def\thefootnote{\fnsymbol{footnote}}
\bigskip

\begin{center}
{\fontsize{20}{15}\selectfont  \bf Beyond Schr\"{o}dinger-Poisson: Nonrelativistic \\ 
\vspace{2mm} Effective Field Theory for Scalar Dark Matter}
\\[0.5cm]
\end{center}
\vspace{0.2cm}

\begin{center}
{Borna Salehian$^{1}\blfootnote{salehian@ipm.ir}$, Hong-Yi Zhang$^{2}\blfootnote{hongyi@rice.edu}$,\\ Mustafa A.~Amin$^{2}\blfootnote{mustafa.a.amin@rice.edu}$, David I.~Kaiser$^{3}\blfootnote{dikaiser@mit.edu}$, Mohammad Hossein Namjoo$^{1}\blfootnote{mh.namjoo@ipm.ir}$}
\\[.7cm]

{\small \textit{$^{1}$School of Astronomy, 
		Institute for Research in Fundamental Sciences (IPM), Tehran, Iran}}
	
\vspace{0.5cm}
{\small \textit{$^{2}$Department of Physics and Astronomy, Rice University, Houston, U.S.A.}} \\ 

\vspace{0.5cm}
{\small \textit{$^{3}$Department of Physics, Massachusetts Institute of Technology, Cambridge, U.S.A.}} \\ 

\end{center}

\vspace{1cm}


\hrule \vspace{0.5cm}
{\small  \noindent \textbf{Abstract} \\[0.2cm]
\noindent
Massive scalar fields provide excellent dark matter candidates, whose dynamics are often explored analytically and numerically using nonrelativistic Schr\"{o}dinger-Poisson (SP) equations in a cosmological context. In this paper, starting from the nonlinear and fully relativistic Klein-Gordon-Einstein (KGE) equations in an expanding universe, we provide a systematic framework for deriving the SP equations, as well as relativistic corrections to them, by integrating out `fast  modes' and including nonlinear metric and matter contributions. We provide explicit equations for the leading-order relativistic corrections, which  provide insight into deviations from the SP equations as the system approaches the relativistic regime. Upon including the leading-order corrections, our equations are applicable beyond the domain of validity of the SP system, and are simpler to use than the full KGE case in some contexts. As a concrete application, we calculate the mass-radius relationship of solitons in scalar dark matter and accurately capture the deviations of this relationship from the SP system towards the KGE one.    
\vspace{0.5cm} \hrule}

\end{titlepage}

\hrule

\tableofcontents

\vspace{5mm}

\hrule

\vspace{5mm}
\section{Introduction}
Axions and axion-like particles are well motivated candidates for dark matter \cite{Wilczek:1977pj,Peccei:1977hh,Preskill:1982cy,Abbott:1982af,Dine:1982ah,Matos:1999et,Ringwald:2014vqa,Hu:2000ke,Hui:2016ltb,Arvanitaki:2009fg}. In cosmological and astrophysical contexts, the typical occupation number of the axion-like fields is large (for masses $m_a\ll\mathcal{O}[10] \rm eV$), allowing for a classical field description of the dynamics \cite{Guth:2014hsa,Hertzberg:2016tal}. The classical field/wave dynamics are manifest in effects on the scale of the de Broglie wavelength of the particles. Since the mass of such axion-like fields is typically taken to be $m_a\lesssim 10^{-5}\rm eV$, the effects of wave dynamics can occur on macroscopic or even astrophysical scales, giving rise to the possibility of distinguishing such dark matter from other dark matter candidates. Novel dynamics such as the formation of solitons \cite{Kolb:1993hw, Schive:2014dra, Levkov:2018kau, Amin:2019ums}, interference patterns \cite{Schive:2014dra}, transient vortices \cite{Hui:2020hbq}, and suppression of power below the de Broglie length-scale of the particles \cite{Hu:2000ke,Urena-Lopez:2015gur,Hlozek:2014lca} is quite generic in axion-like dark matter. For recent reviews, see Refs.~\cite{Marsh:2015xka,Ferreira:2020fam, Hui:2021tkt}.

The axion field oscillates rapidly on the time-scale of order $m_a^{-1}$, whereas its spatial variations are on length-scales $L\sim (v m_a)^{-1}$, where $v\ll 1$ is the typical velocity of the axion particles. Moreover $m_a/H\gg 1$ (where $H$ is the Hubble parameter) within a few e-folds of expansion after the axion field starts oscillating. Together, these considerations indicate that a nonrelativistic description of the field, obtained by integrating out the rapid variations (in time), might be possible and fruitful for cosmological and astrophysical applications. Such an effective nonrelativistic theory would be extremely useful (both analytically and computationally), since one would no longer need to resolve the rapid oscillations of the field.

In the present paper, we start from the relativistic Lagrangian of a classical, real-valued scalar field within general relativity. By systematically integrating out relativistic degrees of freedom we obtain an effective nonrelativistic description for the system. Our specific approach was first incorporated in Ref.~\cite{Namjoo:2017nia} to obtain an effective field theory (EFT) in Minkowski spacetime for a self-interacting scalar field. It was then generalized for curved spacetimes in Ref.~\cite{Salehian:2020bon}, and more specifically applied to the case of a spatially flat Friedmann-Lema\^{i}tre-Robertson-Walker (FLRW) universe, with the analysis restricted to linearized perturbations.\footnote{Note that the terminology of effective field theory refers to two different approaches. One approach is bottom-up, in which all relevant operators that are consistent with the symmetries are included and then the coefficients are fixed by matching with experiments. This approach is incorporated for example in the EFT of inflation \cite{Cheung:2007st} and large-scale structure formation \cite{Carrasco:2012cv}. In contrast, our approach here is top-down, in which an EFT is obtained by taking the low-energy limit of a more complete theory. In this case, the coefficients appearing in the EFT are fixed by the parameters given in the more complete theory. This approach has been used for axion dark matter, for example in Refs.~\cite{Namjoo:2017nia,Eby:2018ufi,Braaten:2018lmj}. Useful comparisons of the different top-down results are also provided in the same papers.} However, one important feature of dark matter is its ability to form dense, nonlinear structures due to gravitational instability in an expanding universe. The focus of this work is therefore to develop an EFT without any assumptions regarding the amplitude of the density perturbations of dark matter within an expanding universe. In this sense we obtain an EFT for axion dark matter in the nonlinear regime. Although metric perturbations are expected to remain small (at least in typical cosmological contexts \cite{Adamek:2013wja,Adamek:2015eda}), we systematically go beyond linear order in the metric perturbations as well. 

\begin{figure}[t] 
   \centering
   \includegraphics[width=6.5in]{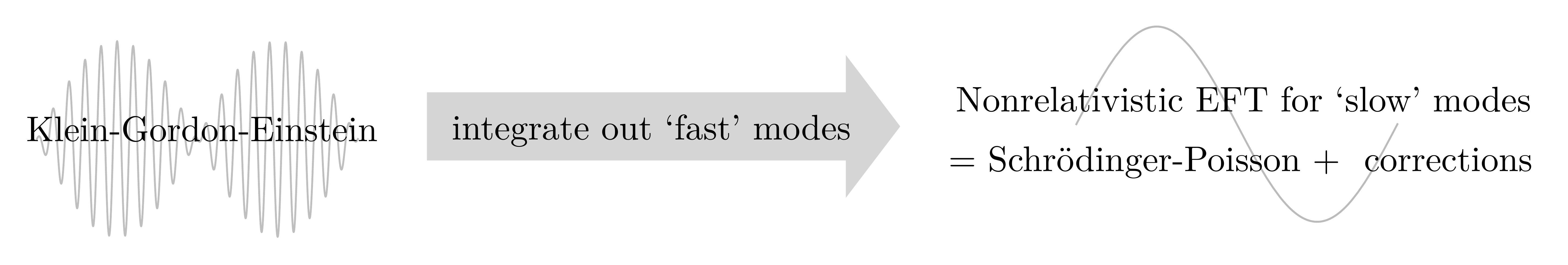} 
\caption{Schematic approach of our EFT method for identifying systematic corrections to the Schr\"{o}dinger-Poisson equations.  }
   \label{fig:schematic}
\end{figure}

The leading-order result in our EFT is consistent with the Schr\"{o}dinger-Poisson (SP) system in an expanding universe, which is widely used in the literature \cite{Hu:2000ke}. For example, the SP system has enabled long-time-scale simulations of nonlinear structure formation of axion-like fields \cite{Schive:2014dra,Mocz:2019pyf,Mocz:2019uyd}. It has also been used to understand the cosmological formation, gravitational clustering, and scattering of solitons with strong self-interactions in the early and contemporary universe \cite{Amin:2019ums}. Mirroring the late-universe simulations, purely gravitational growth of structure in the very early universe was pursued in Ref.~\cite{Musoke:2019ima} with the help of the SP system. The SP system was used for numerically exploring mergers and collisions of solitons with and without self-interactions in axion-like dark matter \cite{Schwabe:2016rze,Glennon:2020dxs}, along with their non-gravitational consequences \cite{Hertzberg:2020dbk,Amin:2020vja}. The SP system was at the heart of exploring dynamical friction \cite{Lancaster:2019mde}, relaxation \cite{Bar-Or:2018pxz}, turbulence \cite{Mocz:2017wlg}, halo substructure \cite{Du:2016zcv,May:2021wwp}, kinetic nucleation of solitons \cite{Levkov:2018kau,Kirkpatrick:2020fwd}, and the dynamics of transient vortices in fuzzy dark matter scenarios \cite{Hui:2020hbq}. A number of existing numerical algorithms and codes are being used to explore the nonlinear dynamics of the SP system. (See, e.g., Refs.~\cite{Veltmaat:2016rxo,Mocz:2017wlg, Edwards:2018ccc}.)

Given its importance and widespread use, it is critical to understand the domain of validity of the SP system as well as expected deviations from it. With our systematic expansion, which relies upon integrating out the dynamics on short time-scales, we go beyond the leading-order SP system of equations and capture quantitative deviations expected due to relativistic corrections. See Fig.~\ref{fig:schematic}. These deviations are expected to be small in most cosmological contexts in the late universe, when the fields are essentially nonrelativistic. Nevertheless, explicit expressions for the relativistic corrections to the Schr\"{o}dinger-Poisson system can pinpoint which particular physical attribute of the system dominates the corrections. For example, one can hope to understand the relative importance of large gradients in the field, deviations from Newtonian gravity, and self-interactions of the scalar field, and at what order in the relativistic corrections vector and tensor perturbations of the metric become relevant as one moves beyond the Schr\"{o}dinger-Poisson system. This understanding, in turn, can clarify the domain of validity of the Schr\"{o}dinger-Poisson system, and provide insights into the most efficient way of including relativistic corrections in different physical contexts (such as those discussed in the previous paragraph). The corrections can also point the way towards exploring deviations from general relativity or characterizing the type of field content making up the dark matter. Furthermore, they might point to symmetries in the problem that are lost or restored as we go beyond the SP system.

As an explicit application of our nonrelativistic EFT equations, we explore the mass-radius relation for dense solitons in the axion field. We demonstrate that our EFT better approximates the fully relativistic solution within the mildly relativistic regime compared to the SP equations alone.  Although our EFT with leading-order relativistic corrections is more complicated than the Schr\"{o}dinger-Poisson system, it is still easier to use numerically and analytically than the fully relativistic Klein-Gordon-Einstein (KGE) equations.

The rest of the paper is organized as following. In Section \ref{sec2}, we provide the fundamental equations and model for a real-valued scalar field within general relativity. We define the small parameters and state the relevant approximations which allow us to simplify the general system. In Section \ref{eftnr} we define and apply our procedure to systematically remove the relativistic modes from our system, and arrive at the nonrelativistic EFT. The main results of our paper are included in this section. In Section \ref{sec:spherical}, we use our EFT equations to improve upon the mass-radius relationship of solitons obtained from the SP system. We end with discussion and conclusions in Section \ref{sec:summary}. A number of appendices provide technical details of calculations to which we allude but do not explicitly provide within the main text.

\section{Setting the stage}\label{sec2}

In this paper we consider the nonlinear and inhomogeneous dynamics of a scalar field. Having in mind cosmological applications, we consider an expanding universe which contains a perfect fluid in addition to the scalar field; the additional fluid contributes to the homogeneous and isotropic background. However, we neglect perturbations of the perfect fluid and assume that gravitational collapse is only sourced by the scalar field. Our theory therefore takes the following form:
\eq{
	\label{action}
	S=\int\dd[4]{x}\sqrt{-g}\left[\frac{1}{2}\Mpl^2R+\mathcal{L}_\vp\right]+\text{Background fluid}\,,
}
where $\Mpl \equiv 1 / \sqrt{8 \pi G_N} = 2.43 \times 10^{18}$ GeV is the reduced Planck mass. The scalar field and the background fluid are described by a Lagrangian density (${\cal L}_\varphi$) and an energy-momentum tensor ($T_f$), respectively,
\ba 
\label{lag}
\mathcal{L}_\vp&=&-\left[\frac{1}{2}g^{\mu\nu}\p_\mu\vp\p_\nu\vp+\frac{1}{2}m^2\vp^2+V_{\rm int}\right]\,,\qquad V_{\rm int}=\frac{1}{4!}\lambda\vp^4+\frac{1}{6!}\frac{\kappa}{\Lambda^2}\vp^6+\dots\,,
\\\nonumber\\
(T_f)^\mu{}_\nu&=&p\,\delta^\mu{}_\nu+(p+\rho)u^\mu u_\nu.
\ea 
where $m$ is the mass of the scalar field, $\lambda$ and $\kappa$ are dimensionless coupling constants, and $\Lambda$ is a large mass scale (compared to $m$). Note that we do not necessarily assume that the scalar field is the axion. However, for the specific case of an axion field, nonperturbative effects generate a periodic potential which is usually approximated by 
\eq{
V(\vp)=m^2f_{\rm a}^2\left[1-\cos(\frac{\vp}{f_{\rm a}})\right]\approx\frac{1}{2}m^2\vp^2-\frac{1}{4!}\left(\frac{m^2}{f_{\rm a}^2}\right)\vp^4+\frac{1}{6!}\left(\frac{m^2}{f_{\rm a}^4}\right)\vp^6+\dots\,,
}
where $f_{\rm a}$ is the axion decay constant and the last approximate equality holds for small field values (compared to $f_{\rm a}$), which is a good approximation in the nonrelativistic limit. As a result, for the axion field, the parameters defined in Eq.~\eqref{lag} are not independent; namely, we have $\lambda=-\kappa=-m^2/f_{\rm a}^2$ and $\Lambda=f_{\rm a}$. In a more general scalar effective field theory, the parameters $\lambda$ and $\kappa$ are independent, but the $\kappa$ term is expected to be suppressed compared to the $\lambda$ term for a sufficiently large cutoff $\Lambda$. In what follows, we will assume such a hierarchy but our approach for obtaining the nonrelativistic EFT can easily be extended to a situation with no hierarchy.

As discussed in the introduction, in the nonrelativistic regime the dynamics of the scalar field are dominated by oscillations with frequency almost equal to its mass $m$. Thus it is reasonable to rewrite the scalar field in terms of a new, complex variable $\psi$ (the so-called ``nonrelativistic field") by
\eq{
\label{redef1}
\vp(t,\vb{x})= \frac{1}{\sqrt{2m}}\left[ e^{-imt}\psi(t,\vb{x})+e^{imt}\psi^*(t,\vb{x})\right]\,.
}
The remaining time or space dependence, encoded in $\psi$, is expected to be dominated by low-energy physics (i.e., lower than the mass scale), so that $\psi$ is a slowly varying function of time and space (compared to the dominant frequency of the system given by $m$). However, due to the nonlinearities involved in the system, high-frequency oscillations appear in $\psi$ with small amplitudes. The task of the following section is to integrate out such high-frequency modes and obtain an effective theory for the slowly varying mode.    

It should be noted that the field redefinition of Eq.~\eqref{redef1} is not a one-to-one correspondence between the real field $\vp$ and the complex nonrelativistic field $\psi$. This fact, which is usually overlooked in the literature, was first addressed in Ref.~\cite{Namjoo:2017nia} for Minkowski spacetime. In that work, the authors assumed a relation similar to Eq.~\eqref{redef1} as a transformation in phase space with an accompanying redefinition for the conjugate momentum, which together make the transformation canonical and invertible. A nonlocal operator was also introduced in Ref.~\cite{Namjoo:2017nia}, which simplifies the derivation of the EFT in Minkowski spacetime. However, as discussed in Ref.~\cite{Salehian:2020bon}, this strategy is not very helpful for the case of curved spacetimes. An alternative approach is to remove the redundancy in Eq.~\eqref{redef1} by adding a constraint on the nonrelativistic field $\psi(t,\vb{x})$. One convenient choice for the constraint is \cite{Salehian:2020bon}
\eq{
\label{cons}
e^{-imt}\dot{\psi}+e^{imt}\dot{\psi}^*=0\,,
}    
where the overdot denotes a time derivative. This constraint implies an equation of motion for $\psi$ that is first order in time derivatives. By using Eq.~\eqref{cons} one can show that
\eq{
\label{redef2}
\dot{\vp}(t,\vb{x})=-i\sqrt{\frac{m}{2}}\left[e^{-imt}\psi(t,\vb{x})-e^{imt}\psi^*(t,\vb{x})\right]\,.
} 
It is possible to apply the above field redefinitions at the level of the action and write the Lagrangian of Eq.~\eqref{lag} in terms of $\psi$ and $\psi^*$ (see Ref.~\cite{Salehian:2020bon}). However, we are most interested in the equations of motion in terms of the nonrelativistic field. Applying Eqs.~\eqref{redef1} and \eqref{redef2} to the Klein-Gordon equation yields
\eq{
\label{eq:psidotD}
ig^{00}\dot{\psi}+\mathcal{D}\psi+e^{2imt}\mathcal{D}^*\psi^*=-\frac{e^{imt}}{\sqrt{2m}}\dv{V_{\rm int}}{\vp}{}(\psi,\psi^*)\,,
}
where $\mathcal{D}$ is a differential operator defined by
\eq{
\label{Dorg}
\mathcal{D}=\frac{m}{2}\left(g^{00}+1\right)+\frac{i\,\p_\mu(\sqrt{-g}g^{0\mu})}{2\sqrt{-g}}+\left(ig^{0i}-\frac{\p_\mu(\sqrt{-g}g^{\mu i})}{2m\sqrt{-g}}\right)\p_i-\frac{1}{2m}g^{ij}\p_i\p_j\,,
}
and $V_{\rm int}$ is given in Eq.~\eqref{lag}, which here is written in terms of $\psi$ and $\psi^*$ using Eq.~\eqref{redef1}. Eq.~\eqref{eq:psidotD} is a generalized Schr\"odinger equation in an arbitrary spacetime. Notice that Eq.~\eqref{eq:psidotD} is exact and there exists a one-to-one map from the complex field $\psi$ to the real field $\varphi$ and its conjugate momentum. Also note the appearance of rapidly oscillating factors. A common approximation is to neglect such terms, under the assumption that they average to zero. Whereas this is true at leading order (which leads to the SP equations for the scalar dark matter, as we discuss below), these terms will play a crucial role in the derivation of our EFT, as shall be discussed in Sec.~\ref{eftnr}. Note that the oscillatory terms are present even in a free theory with $V_{\rm int}=0$, which then leads to a tower of higher spatial-derivative terms in the free EFT; these terms can be thought of as the expansion of the relativistic energy in the large-mass limit \cite{Namjoo:2017nia}. 

To fully describe the system, the Schr\"odinger equation of Eq.~\eqref{eq:psidotD} must be accompanied by the Einstein field equations as well as the energy-momentum conservation for the fluid,
\eq{
G^\mu{}_\nu=\frac{1}{\Mpl^2}\big[T^\mu{}_\nu+(T_f)^\mu{}_\nu\big]\,,\qquad (T_f)^\mu{}_{\nu;\mu}=0\,.
}
Let us emphasize again that in what follows, for simplicity, we will ignore perturbations of the background fluid. 

\subsection{FLRW metric with perturbations}
\label{sec:FLRW}
Since we have in mind the application of our EFT to cosmology, we consider a perturbed expanding universe as a specific case of the general discussion in the previous section. As noted above, we intend to study the inhomogeneities in the scalar field $\psi (t, {\bf x})$ nonlinearly. This implies that, contrary to the case for linear perturbation theory, the vector and tensor modes of the spacetime metric may play a nontrivial role in the dynamics of the scalar degrees of freedom. As a result, here we start from a general metric, including all forms of the metric perturbations, and then estimate the contribution of each type of mode to the dynamics of the scalar field. It is convenient to work with the ADM metric decomposition, which is given by
\eq{
\label{metric1}
\dd{s}^2=-N^2\dd{t}^2+\gamma_{ij}(\dd{x^i}+N^i\dd{t})(\dd{x^j}+N^j\dd{t})\,,
}
where $N$, $N^i$ and $\gamma_{ij}$ are the lapse function, shift vector, and the first fundamental form, respectively. We remove the gauge redundancy by the following choice of the metric components:
\eq{
N=e^\Phi\,,\qquad N^i=\frac{1}{a}\sigma^i\,,\qquad\gamma_{ij}=a^2e^{-2\Psi}(e^h)_{ij}\,,
} 
where we have
\eq{
\label{eqTT}
\p_i\sigma^i=\delta^{ij}h_{ij}=\p_ih^i{}_j=0\,,
}
and we lower and raise the Latin indices with $\delta_{ij}$ and $\delta^{ij}$. The background geometry is FLRW spacetime and $a(t)$ is its corresponding scale factor. We have fixed the gauge by requiring the scalar mode of $g_{0i}$ and the vector and some of the scalar modes of $g_{ij}$ to vanish. This  choice of gauge can be retained to all orders of perturbations and is a natural generalization of the Newtonian gauge, which is particularly convenient for the system in the nonrelativistic regime. Note that we think of the above metric as perturbative in $\Phi$, $\Psi$, $\sigma^i$ and $h_{ij}$ (while we treat the scalar field $\psi$ nonperturbatively), which we will justify shortly.  From Eqs.~\eqref{metric1}-\eqref{eqTT}, one has 
\eq{
\label{metric2}
\spl{
\sqrt{-g}=N\sqrt{\gamma}=a^3e^{\Phi-3\Psi}\,,\qquad g^{00}&=\frac{-1}{N^2}=-e^{-2\Phi}\,,\qquad g^{0i}=\frac{N^i}{N^2}=\frac{1}{a}e^{-2\Phi}\sigma^i\,,\\ g^{ij}=\gamma^{ij}-\frac{1}{N^2}N^iN^j&=\frac{1}{a^2}\Big[e^{2\Psi}(e^{-h})^{ij}-e^{-2\Phi}\sigma^i\sigma^j\Big]\,,\qquad 
}}    
which may be used in Eq.~\eqref{eq:psidotD} as well as for the Einstein field equations. 

\subsection{Power counting}
\label{sec:power-counting}
As we take the nonrelativistic limit of a relativistic theory, several small parameters/operators appear, which allows us to organize different terms that arise in the EFT. In this section we shall identify these small parameters.  Furthermore, in the nonrelativistic limit and as a result of the source of gravitational perturbations being a scalar field, there exists a hierarchy among the amplitudes of the scalar, vector, and tensor modes of the perturbed spacetime metric. As we will see below, the scalar modes dominate and the amplitude of the vector mode is larger than that of the tensor modes. 

Let us start by looking at the Klein-Gordon equation (ignoring metric perturbations):
\eq{
\label{eqphi}
\spl{
&\ddot{\vp}+m^2\vp\\
&+3H\dot{\vp}-\frac{\nabla^2\vp}{a^2}+\frac{1}{3!}\lambda\vp^3+\frac{1}{5!}\frac{\kappa}{\Lambda^2}\vp^5+\dots=0\,.
}
}
As stated above, in the nonrelativistic regime the mass term is the dominant contribution to the time evolution of the scalar field, and all other terms are suppressed. We have written the equation of motion in Eq.~\eqref{eqphi} in two different lines to emphasize this hierarchy. Demanding that the terms on the second line are smaller than those on the first line, we identify the following small parameters in the nonrelativistic limit:
\eq{
\epsilon_{\scriptscriptstyle H}\sim\frac{H}{m}\ll1\,,\qquad	
\epsilon_x\sim\Bigg|\frac{\nabla^2}{m^2a^2}\Bigg|\ll1\,,\qquad\epsilon_\lambda\sim |\lambda|\frac{\vp^2}{m^2}\ll1\,,\qquad\epsilon_\kappa\sim|\kappa|\frac{\vp^4}{m^2\Lambda^2}\ll1\,.
} 
The first parameter quantifies the smallness of the expansion rate compared to the mass scale. In the opposite regime, when the Hubble scale is larger than $m$, the scalar field does not oscillate and cannot mimic dark matter behavior. The second parameter quantifies the smallness of the typical momentum of the dark matter ``particles" compared to $m$, while the last two parameters specify the smallness of self-interactions. Note that if we assume that $\lambda$ and $\kappa$ are of the same order and $\Lambda$ is a very large mass scale, then one can see that $\epsilon_\kappa$ is much smaller than $\epsilon_\lambda$. In fact for the special case of the axion-like field we have $\epsilon_\kappa=\epsilon_\lambda^2$. Although we do not restrict ourselves to the axion, we do assume a similar hierarchy between these two parameters, with $\epsilon_\kappa\sim {\cal O} (\epsilon_\lambda^2)$. 

Next we study the hierarchy among the dynamical variables, which follows as a consequence of the system being nonrelativistic. For these estimates, one can use the Einstein field equations. However, most of the approximate relations can also be estimated by considering symmetries and other simple relationships. First we note that in order for the system to remain nonrelativistic even amid the gravitational dynamics, the gravitational potentials must remain small,  
\eq{
\epsilon_g\equiv |\Phi|\,\sim|\Psi|\ll1\,.
} 
The fact that $\Phi$ and $\Psi$ are expected to be of the same order in $\epsilon_g$ is discussed below. From the 00 component of the Einstein field equations one can obtain the Poisson-like equation for $\Psi$,  leading to
\eq{
\label{phi}
\frac{\nabla^2\Psi}{a^2}\sim\frac{m^2\vp^2}{\Mpl^2}\,.
}
Further, the Poisson equation implies that there is another small parameter related to the amplitude of the scalar field, 
\eq{
\label{ephi}
\epsilon_\vp\equiv \frac{|\vp|}{\Mpl}\sim\frac{|\psi|}{\Mpl\sqrt{m}}\ll1\,,
} 
from which we find $\epsilon_\vp^2\sim\epsilon_x\epsilon_g$. In addition, by using the trace-free part of the $ij$-component of the Einstein field equations we can see that
\eq{
	\label{phi-psi-diff}
\nabla^2(\Phi-\Psi)\sim\frac{(\nabla\vp)^2}{\Mpl^2}\,,\quad\implies\quad\Phi-\Psi\sim\epsilon_\vp^2\, , 
}
that is, the difference between the two gravitational potentials is typically one order smaller than the gravitational potentials themselves. 
Further, from the $0i$-component of the Einstein field equations we find
\eq{
\label{sigma1}
\frac{1}{a}\nabla^2\sigma_i\sim\frac{1}{\Mpl^2}\dot{\vp}\p_i\vp\,,\quad\implies\quad	\sigma_j\sim\epsilon_x^{1/2}\epsilon_g\,,
} 
where we have used $\dot{\vp}\sim m\vp$. Note that the relation between the vector mode and the scalar field (which acts as its source) could also be identified from symmetries and dimensional analysis. Finally, by using the $ij$-component of the Einstein field equations (or, again, by  symmetries), we find 
\eq{
	\label{hij1}
\nabla^2h_{ij}\sim\frac{\p_i\vp\p_j\vp}{\Mpl^2}\,,\quad\implies\quad h_{ij}\sim\epsilon_\vp^2\,.
}
In the following, instead of keep tracking of these small parameters individually, we collectively denote all of them by $\epsilon=\{\epsilon_{\scriptscriptstyle H},\epsilon_x,\epsilon_\lambda,\epsilon_g,\epsilon_\vp\}$ and work up to appropriate order in $\epsilon$. 
This effectively means that we assume all small parameters are of the same order (except for $\epsilon_\kappa$, which is one order smaller). Depending on the application, it is expected that a hierarchy among the small parameters exists, in which case our EFT would be simplified. By using our approach, any higher-order term in the EFT can be derived systematically. However, in the main text of this paper we focus on the leading relativistic corrections to the equations of motion. In other words, we are primarily interested in terms which are one order smaller in $\epsilon$  than the SP system of equations, which constitute the leading relativistic terms. In Appendix \ref{app:NREFT2} we proceed one order beyond the leading-order corrections for the specific case of spherically symmetric solitonic solutions.

\subsection{Scalar and vector equations}
\label{sec:equations}

By using our power-counting arguments, one can obtain a set of equations for the gravitating scalar dark matter in an expanding background at the requisite order in $\epsilon$.
Because our primary interest is the evolution of the scalar modes, we will be dealing with scalar equations of motion; hence the tensor modes $h_{ij}$ cannot appear by themselves, but will always enter the equations of motion accompanied by at least two spatial derivatives. This implies that the tensor modes would only appear at $\order{\epsilon^3}$, according to Eq.~\eqref{hij1}. Similarly, the vector mode $\sigma^i$ appears with at least one spatial derivative which, upon using Eq.~\eqref{sigma1}, implies that it appears at $\order{\epsilon^2}$ in the scalar equations of motion.

Based on the above considerations, the generalized Schr\"{o}dinger equation of Eq.~\eqref{eq:psidotD}, the Einstein field equations that reduce to the Poisson equation in the nonrelativistic limit, and the equation for the vector mode take the following approximate form:
\begin{align}
\label{eq:psidotD3}
&i\dot{\psi}+\tilde{\mathcal{D}}\psi+e^{2imt}\tilde{\mathcal{D}}^*\psi^* =e^{imt}e^{2\Phi}\mathcal{J}+\order{\epsilon^{4}}\,,
\\ \nonumber \\
	\label{eqPhi-simp}
		&	\frac{\nabla^2}{a^2}\Phi+3e^{-2(\Phi+\Psi)}\left(H\dot{\Phi}+2H\dot{\Psi}-\dot{\Phi}\dot{\Psi}-\dot{\Psi}^2+\ddot{\Psi}-\frac{\ddot{a}}{a}\right)
		= \frac{e^{-2\Psi}}{2\Mpl^2}\Big[(\rho+3p)+\mathcal{S}_\Phi\Big]+\order{\epsilon^{4}}\,,
	\\\nonumber \\
		\label{eqPsi-simp}
&	\frac{\nabla^2\Psi}{a^2}-\frac{(\nabla\Phi)^2}{2a^2}+\frac{3}{2}e^{-2(\Phi+\Psi)}\left(H^2+\dot{\Psi}^2-2H\dot{\Psi}\right)= \frac{	e^{-2\Psi}}{2\Mpl^2}\Big[\rho+\mathcal{S}_\Psi\Big]+\order{\epsilon^{4}}\,,
	\\ \nonumber \\
		\label{eqsigma-simp}
&	\frac{\nabla^4}{a}\sigma_i =\frac{2i}{\Mpl^2}\Big[ \left( \nabla^2\psi+(\nabla\psi \cdot \nabla) \right) \p_j\psi^*- \left( \nabla^2\psi^*+(\nabla\psi^* \cdot \nabla) \right) \p_j\psi\Big]+\order{\epsilon^{9/2}}\, ,
\end{align}
where we have defined
\eqa{
	\label{D-simp}
\tilde{\mathcal{D}}&=\frac{m}{2}\left(1-e^{2\Phi} \right)+\frac{e^{4\Phi}}{2ma^2}\nabla^2+\frac{i}{2}\big(3H-\dot{\Phi}-3\dot{\Psi}\big)-\frac{i}{a}\vec{\sigma} \cdot \nabla\,,\\
	\label{J-simp}
	\mathcal{J}&=\frac{1}{3!}\lambda\vp^3+\frac{1}{5!}\frac{\kappa}{\Lambda^2}\vp^5\, , \qquad 
	\mathcal{S}_\Phi = 2e^{-2\Phi}\dot \vp^2 -m^2\vp^2 -\dfrac{2\lambda}{4!}\vp^4,
	\\
	\label{Spsi-simp}
	\mathcal{S}_\Psi   &= \dfrac12 e^{-2\Phi}\dot \vp^2 +\dfrac{e^{2\Phi}}{2a^2}(\nabla \vp)^2+\dfrac12 m^2\vp^2 +\dfrac{\lambda}{4!}\vp^4 \, .
}
Within the expressions in Eqs.~(\ref{D-simp})-(\ref{Spsi-simp}), the fields $\vp$ and $\dot \vp$ need to be replaced by $\psi$ and $\psi^*$ using Eqs.~\eqref{redef1} and \eqref{redef2} (see Appendix~\ref{equations}). Note that we have replaced $\Psi$ with $\Phi$ in several terms, because the difference between the two gravitational potentials is one order smaller than $\Phi$ and $\Psi$ themselves. (See  Appendix~\ref{equations} for equations that are nonperturbative in $\Phi$ and $\Psi$, though still linear in the vector and tensor modes.) To avoid clutter we did not expand the exponential factors, but one needs to keep in mind that they are only relevant to appropriate order in their Taylor expansion. Notice that since --- at this stage --- different variables may contain highly oscillating contributions, it is not necessarily the case that the time derivative operator is small. Moreover, as a result of the assumed hierarchy between the self-interaction terms (i.e., $\epsilon_\kappa \sim \epsilon_\lambda^2$), the $\kappa$ term only appears in the Schr\"odinger equation at the current working  order. 

In general, the order of terms that are neglected must be compared with the leading-order terms. For example, in the Schr\"odinger equation of Eq.~\eqref{eq:psidotD3}, the leading-order terms (such as $\nabla^2 \psi/m^2a^2$) are already of $\order{\epsilon^2}$, according to our power counting. This implies that we are neglecting some terms that are at least two orders higher in $\epsilon$. It is thus evident that we have only kept the leading-order nontrivial corrections, which is indeed the case for all other equations. To go beyond that, one needs a more accurate set of equations. Toward that end, in Appendix~\ref{equations} we consider the same set of equations but treat the scalar gravitational potentials nonperturbatively. Then we may apply our EFT beyond leading-order corrections for specific scenarios, such as spherically symmetric solitonic solutions (see Appendix~\ref{app:NREFT2}). In such cases, in which one may appeal to additional symmetries, deriving the next-higher-order corrections can be simplified.  Throughout Sec.~\ref{eftnr}, however, we develop an EFT applicable to more general situations.

We shall see in Sec.~\ref{eftnr} that one can obtain the SP system of equations from Eqs.~\eqref{eq:psidotD3} and \eqref{eqPhi-simp} to leading order in the EFT, while corrections appear at the next order. As a final remark, note that at the background level the above set of equations reduce to 
\eqa{
&i\dot{\bar{\psi}}+\frac{3i}{2}H \left( \bar{\psi}-e^{2imt}\bar{\psi}^* \right)=e^{imt}\mathcal{\bar{J}}, \, \quad 
3\Mpl^2H^2=\rho+\bar{\mathcal{S}}_\Psi, \, \quad 
	\frac{\ddot{a}}{a}=-\frac{1}{6\Mpl^2}\left[(\rho+3p)+\bar{\mathcal{S}}_\Phi\right], \quad 
\nonumber 	\\ 
\label{bg}
&\dot{\rho}+3H(\rho+p)=0\,,
} 
where an overbar indicates that the quantity is evaluated at the spatially homogeneous background level (see Ref.~\cite{Salehian:2020bon}). In Eq.~(\ref{bg}) we have also included the continuity equation for the additional perfect fluid, which is assumed to be spatially homogeneous.

So far, by removing unnecessary terms, we have already taken the first step toward identifying the leading-order corrections in the nonrelativistic EFT. In principle, one can solve Eqs.~(\ref{eq:psidotD3})-(\ref{eqsigma-simp}) numerically to obtain the dynamics of the scalar field. However, such numerical computation is, in general, a difficult task due to the rapidly oscillating factors appearing in those equations. In Sec.~\ref{eftnr} we shall remove such factors in a systematic way (instead of naively neglecting them) and obtain their corresponding corrections to the slowly varying variables. We will see that this procedure leads to nontrivial corrections at a given working order in $\epsilon$, and hence cannot be neglected.

\section{Effective field theory in the nonrelativistic limit}\label{eftnr}

As stated in the previous section, we are interested in the slowly varying modes of dynamical variables. However, due to the appearance of oscillatory factors in the equations of motion, the dynamics of slowly varying quantities will be affected by rapidly oscillating terms. The situation is illustrated in Fig.~\ref{fig:modes}, which shows the typical behavior of the variables in the frequency domain (i.e., the Fourier transform of time-dependent variables). As shown in Fig.~\ref{fig:modes}, the zero mode (which translates to the slowly varying mode in the time domain) dominates,   although modes with nonzero frequencies --- close to integer multiples of the mass $m$ --- also exist in the spectrum, albeit with subdominant amplitudes.

To obtain a theory entirely in terms of  slowly varying quantities, we must {\it integrate out} modes associated with these rapid oscillations. This is a nontrivial task because the modes associated with rapid oscillations are sourced by the slowly varying mode, and they in turn backreact on the evolution of the slowly varying mode, affecting its dynamics.

Working in the time domain, we may apply a \emph{smearing operator} to each variable in order to extract the slowly varying part of the variables. That is, we may take a time average of each variable with a suitable choice of window function \cite{Salehian:2020bon}: 
\fg{
\centering
\includegraphics[width=0.9\textwidth]{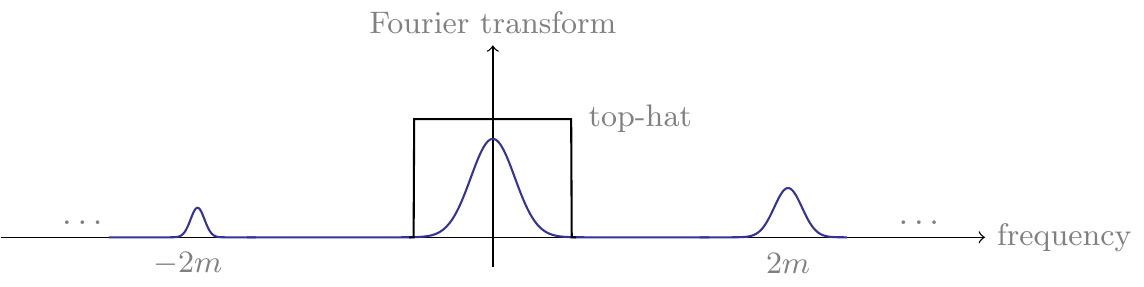} 
\caption{Typical frequency  spectrum of the variables in the problem.  The system is dominated by the slowly varying mode, though modes associated with rapid oscillations also arise in the spectrum. By applying the smearing operator of Eq.~(\ref{eq:smearing}), we may extract the slowly varying part.}\label{fig:modes}
}
\eq{
\label{eq:smearing}
X_s=\ev{X}\equiv\int\dd{t'}W(t-t')X(t')\,,
}
in which $W(t)$ is the window function and $X_s$ is the ``slow mode" of the variable $X$. In Ref.~\cite{Salehian:2020bon} it has been shown that the top-hat window function in the frequency domain, which becomes $\rm{sinc}(t)$ in  the time domain, is a suitable choice. Besides the slow mode $X_s$, each variable also contains a tower of modes  associated with rapid oscillations. Quite generally, one has
\eq{
\label{eq:modexp}
X=\sum_{\nu=-\infty}^{\infty}X_\nu\, e^{i\nu m t}\,,\qquad X_\nu=\ev{X\,e^{-i\nu m t}}\,,
} 
where the coefficients $X_\nu$ depend on both time and space.  We define the ``slow mode" as $X_s =X_{\nu = 0}$, and refer to the modes associated with rapid oscillations as ``nonzero modes," that is, modes $X_\nu$ with $\nu \neq 0$. According to the definition in Eq.~\eqref{eq:modexp}, we have $(X^*)_\nu=X_{-\nu}^*$, and therefore if $X$ is real-valued then the modes $X_\nu$ obey the constraint $X_\nu^*=X_{-\nu}$. 

Note first that the expansion in Eq.~(\ref{eq:modexp}) is exact, as a result of the appropriate  choice of the window function. (See Ref.~\cite{Salehian:2020bon} for an outline of the proof.) Second, we emphasize that the coefficients $X_\nu$ (even with $\nu \neq 0$) are themselves slowly varying functions of time (compared to the frequency of the oscillations) since, as noted in the second expression in Eq.~(\ref{eq:modexp}), the $X_\nu$ may be represented by the smearing operator acting on $X$ (weighted by an appropriate phase). In fact, as shall be made explicit below, the $\nu\ne 0$ modes can be written in terms of the slow mode $X_s$ (since, to reiterate, they are sourced by the slow mode). We may therefore identify a new small operator, namely, the time-derivative operator acting on the slow modes,
\eq{
	\epsilon_t \sim \left\vert \frac{\dot{X_s}}{mX_s} \right\vert \ll 1\, ,
}
where $X$ can be any of our variables after the field redefinition, including $\psi$, $\Phi$, $\Psi$, $\sigma_i$, $h_{ij}$, and $a$; the subscript indicates that the slow mode (with $\nu = 0$) is considered.\footnote{Because the functions $X_\nu$ with $\nu \neq 0$ are also slowly varying and can be expressed in terms of $X_s$, we also have $\epsilon_t \sim \vert \dot X_\nu/mX_\nu \vert$ -- at least for $\nu$ not too large.} Thus, we can include the time-derivative operator within the set of small parameters/operators identified below Eq.~(\ref{hij1}), and our EFT will be an expansion in powers of $\epsilon=\{\epsilon_t,\epsilon_{\scriptscriptstyle H},\epsilon_x,\epsilon_\lambda,\epsilon_g,\epsilon_\vp\}$. Note that these small parameters are not all independent; one may derive relations among them by using the equations of motion. Let us emphasize that $\epsilon_t$ is defined as the operator that acts on modes $X_\nu$, rather than on the full fields. In the latter case --- according to the definition in Eq.~\eqref{eq:modexp} --- the time derivative would act on the oscillatory factors, which are not slowly varying, which is why, for example, $\dot{\Psi}$ in Eq.~\eqref{eqPhi-simp} should not be considered as contributing to $\order{\epsilon^2}$ in the power counting.

We will be interested in the effective equations for the slow modes $X_s = X_{\nu = 0}$; therefore we will systematically remove nonzero modes $X_\nu$ with $\nu\neq 0$ from the theory. For a nonrelativistic system, all $\nu\ne0$ modes are suppressed by factors of the small parameters/operators denoted collectively by $\epsilon$. Using power counting to estimate the size of each term that appears in the equations of motion for the nonzero modes, we may solve for them perturbatively in $\epsilon$. To achieve this, we expand the nonzero modes as a power series in $\epsilon$: 
\eq{
\label{pertexp}
X_\nu=\sum_{n=1}^{\infty}X_\nu^{(n)}=X_\nu^{(1)}+X_\nu^{(2)}+\dots\,,\qquad(\nu\neq0)\,.
} 
The superscript $(n)$ denotes the order of magnitude relative to the slow mode: $X^{(n)}_\nu/X_s\sim\order{\epsilon^n}$.\footnote{Note that the power series we use here is slightly different from the one introduced in Ref.~\cite{Salehian:2020bon}. $X^{(n)}_\nu$ here corresponds to what was denoted $X^{(n+1)}_\nu$ in Ref.~\cite{Salehian:2020bon}. } This expansion allows us to solve for the $\nu\ne 0$ modes perturbatively and substitute the solutions back into the equations for the slow mode, resulting in an EFT for the slow modes. This procedure has been done explicitly for an interacting theory in Minkowski spacetime in Ref.~\cite{Namjoo:2017nia} and extended to the case of a linearly perturbed FLRW universe in Ref.~\cite{Salehian:2020bon}. Here we outline essential steps in the derivation, focusing mainly on the Schr\"odinger equation, and discuss additional details in Appendix~\ref{app:EFT_details}.

 Applying the mode expansion of Eq.~\eqref{eq:modexp} to Eq.~\eqref{eq:psidotD3} yields the following equation for each mode:
\eq{
	\label{psinu}
i\dot{\psi}_\nu-\nu m\psi_\nu+\tilde{\mathcal{D}}_\alpha\psi_{\nu-\alpha}+\tilde{\mathcal{D}}^*_{-\alpha}\psi^*_{2+\alpha-\nu}=\left(e^{2\Phi}\right)_\alpha\mathcal{J}_{\nu-\alpha-1}\,,
}    
in which repeated indices are summed over.\footnote{The mode decomposition of equations of motion can be understood as the result of multiplying each equation by $e^{-i \nu m t}$ and then applying the smearing operator. } This equation makes it evident that modes of different $\nu$ couple to each other; in particular, nonzero modes ($\nu\ne 0$) affect the dynamics of the slow mode ($\nu=0$) and the nonzero modes are sourced by the slow mode. Similar results follow from Eqs.~\eqref{eqPhi-simp} for $\Phi$ and \eqref{eqPsi-simp} for $\Psi$ (see Appendix~\ref{app:EFT_details}). We can then solve for the nonzero modes perturbatively. At leading order we find
\eqa{
\label{nonzero}
\psi^{(1)}_\nu&=\left(\frac{\nabla^2}{4m^2a_s^2}-\frac{3iH_s}{4m}-\frac{\lambda|\psi_s|^2}{16m^3}-\frac{1}{2}\Phi_s\right)\psi_s^*\delta_{\nu,2}+\frac{\lambda\psi_s^3}{48m^3}\delta_{\nu,-2}-\frac{\lambda\psi_s^*{}^3}{96m^3}\delta_{\nu,4}\\
\Psi^{(1)}_\nu&=\frac{\psi_s^*{}^2-\bar{\psi}_s^*{}^2}{16m\Mpl^2}\delta_{\nu,2}+\frac{\psi_s^2-\bar{\psi}_s^2}{16m\Mpl^2}\delta_{\nu,-2},\,
\quad 
H^{(1)}_\nu=-\frac{i\bar{\psi}_s^*{}^2}{8\Mpl^2}\delta_{\nu,2}+\frac{i\bar{\psi}_s^2}{8\Mpl^2}\delta_{\nu,-2},
}
where $\psi^{(1)}_\nu$ is derived from Eq.~\eqref{psinu} while $\Psi^{(1)}_\nu$ and $H^{(1)}_\nu$ can be obtained by solving equations after the mode decomposition of Eqs.~\eqref{eqPhi-simp} and \eqref{bg}, respectively. Note that $\delta_{\nu,i}$ is the Kronecker delta function, and the superscript $(1)$ denotes that each term on the right hand side is suppressed by $\order{\epsilon}$ compared to $\psi_s$, $\Psi_s$ or $H_s$ . An expression can also be derived for the $\nu\ne 0$ modes of $\Phi$, with the additional complication that the solution would be nonlocal. Fortunately, among the leading-order corrections (on which we focus in this section), nonzero modes of $\Phi$ do not contribute. Note also that the leading-order nonzero modes of the scale factor vanish. 

We can now use these solutions to replace nonzero modes that appear in the equations for the slow modes. Furthermore, based on the power counting discussed in Sec.~\ref{sec:power-counting}, we can neglect terms that are at higher order compared to the leading-order corrections. After significant algebraic simplification, for the Schr\"odinger equation we find
\bxeq{
\label{eqftkg}
\begin{split}
&i\dot{\psi}_s+\frac{3i}{2}H_s\psi_s+\frac{1}{2ma_s^2}\nabla^2\psi_s-m\Phi_s\psi_s-\frac{\lambda}{8m^2}|\psi_s|^2\psi_s\\
&+\left(\frac{3\rho_s}{8m\Mpl^2}+\frac{|\bar{\psi}_s|^2}{2\Mpl^2}+\frac{|\psi_s|^2}{16\Mpl^2}-\frac{m}{2}\Phi_s^2\right)\psi_s-2i\dot{\Phi}_s\psi_s+\frac{\nabla^4\psi_s}{8m^3a_s^4}\\
&+3\Phi_s\frac{\nabla^2\psi_s}{2ma_s^2}-\frac{\nabla\Phi_s \cdot \nabla\psi_s}{2ma_s^2}-i\frac{\vec{\sigma}_s\cdot\nabla\psi_s}{a_s}+\left(\frac{17\lambda^2}{8m^2}-\frac{\kappa}{\Lambda^2}\right)\frac{|\psi_s|^4\psi_s}{96m^3}\\
&-\frac{\lambda}{16m^4a_s^2}\left(2|\nabla\psi_s|^2\psi_s+\psi_s^2\nabla^2\psi_s^*+2|\psi_s|^2\nabla^2\psi_s+\psi_s^*(\nabla\psi_s)^2\right)=0+\order{\epsilon^4}\,,
\end{split} }
\noindent where the background equations as well as the leading-order Poisson equation are used to simplify the subleading terms. The first line is the standard Schr\"odinger equation in an expanding universe, while all terms on the second line and thereafter are leading-order relativistic corrections. That is, while the first line is $\mathcal{O}(\epsilon^2)$, all terms on the subsequent lines are $\mathcal{O}(\epsilon^3)$.  Note that since $\Psi_s-\Phi_s\sim \mathcal{O}(\epsilon^2)$, we have replaced $\Psi_s$ by $\Phi_s$ if it appears anywhere but the first line. Also note that since $\rho_s\leq 3\Mpl^2 H^2$ we have that $\rho_s/m^2\Mpl^2 \sim \order{\epsilon^2}$ or smaller.

In a similar way, we obtain the effective equation for the gravitational potential $\Phi_s$, starting from Eq.~\eqref{eqPhi-simp}:
\bxeq{
\label{eftPhi}
\begin{split}
&\frac{\nabla^2\Phi_s}{a_s^2}-\frac{m}{2\Mpl^2}(|\psi_s|^2-|\bar{\psi}_s|^2)\\
&+3(3H_s\dot{\Phi}_s+\ddot{\Phi}_s)-\dfrac{1}{\Mpl^2}\left(\rho_s+3p_s+2m|\bar{\psi}_s|^2-(3/2)m|\psi_s|^2\right)\Phi_s\\
&+\frac{3}{8m\Mpl^2a_s^2} \left( \psi_s\nabla^2\psi_s^*+\psi_s^*\nabla^2\psi_s \right) - \frac{\lambda}{8m^2\Mpl^2} \left( |\psi_s|^4-|\bar{\psi}_s|^4 \right)=0+\order{\epsilon^4}\, .
\end{split}}
\noindent Interestingly, notice that the other gravitational potential, $\Psi_s$, decouples from $\Phi_s$ and $\psi_s$ to this order. However,
to close the system of equations, we must add one for the vector modes, which at this order is simply given by Eq.~\eqref{eqsigma-simp} with all variables replaced by their corresponding slow modes:
\bxeq{
\label{eftsigma}
\frac{\nabla^4\vec{\sigma}_s}{a_s}=\frac{2i}{\Mpl^2}\Big[\left( \nabla^2\psi_s+(\nabla\psi_s \cdot \nabla) \right) \nabla\psi_s^* - \left( \nabla^2\psi_s^*+(\nabla\psi_s^* \cdot \nabla) \right)\nabla\psi_s\Big]+\order{\epsilon^{9/2}}\,.
}
\noindent Equations (\ref{eqftkg})-(\ref{eftsigma}) are sufficient for obtaining the leading-order corrections to the Schr\"{o}dinger-Poisson system of equations. However, the gravitational potential $\Psi_s$ might also be of  interest for some purposes, such as lensing effects of compact objects. We therefore present its corresponding effective equation, which can be obtained by a similar procedure, starting from Eq.~\eqref{eqPsi-simp}:
\bxeq{
\label{eftPsi}
\begin{split}
&\frac{\nabla^2\Psi_s}{a_s^2}-\frac{m}{2\Mpl^2}(|\psi_s|^2-|\bar{\psi}_s|^2)\\
&-\frac{(\nabla\Phi_s)^2}{2a_s^2}-3H_s\dot{\Phi}_s-\dfrac{1}{\Mpl^2}\left(\rho_s+2m|\bar{\psi}_s|^2- \frac{3}{2} m |\psi_s|^2\right)\Phi_s
\\
&-\frac{|\nabla\psi_s|^2}{4m\Mpl^2a_s^2}-\frac{\lambda}{32m^2\Mpl^2} \left( |\psi_s|^4-|\bar{\psi}_s|^4 \right) =0+\order{\epsilon^4}.
\end{split}}

Equations~\eqref{eqftkg}-\eqref{eftPsi} are the main results of this paper. It is worth noting that the first lines of Eqs.~\eqref{eqftkg} and \eqref{eftPhi} yield the familiar Schr\"{o}dinger-Poisson system:
\begin{equation}
\begin{aligned}
\label{eq:SP}
&i\dot{\psi}_s+\frac{3i}{2}H_s\psi_s+\frac{1}{2ma_s^2}\nabla^2\psi_s-m\Phi_s\psi_s-\frac{\lambda}{8m^2}|\psi_s|^2\psi_s=0+\order{\epsilon^3}\,,\\
&\frac{\nabla^2\Phi_s}{a_s^2}-\frac{m}{2\Mpl^2} \left( |\psi_s|^2-|\bar{\psi}_s|^2 \right)=0+\order{\epsilon^3}\,.
\end{aligned}
\end{equation}
We do not need Eqs.~\eqref{eftsigma} and \eqref{eftPsi} for the evolution of $\psi_s$ at this order. 

At the FLRW background level, it is easy to check that at leading order there are no corrections to the Friedmann equation or the continuity equation. That is, we have 
\eqa{
\label{eq:SPe}
	3\Mpl^2H_s^2=\rho_s+m|\bar{\psi}_s|^2+\textcolor{black}{\frac{\lambda|\bar{\psi}_s|^4}{16m^2}}+\order{\epsilon^4}, \quad \dot{\rho}_s+3H_s(\rho_s+p_s)=0+\order{\epsilon^4},
}
while the background Schr\"odinger equation receives corrections; these can be obtained by the replacement $\psi_s \to \bar \psi_s$ in Eq.~\eqref{eqftkg} and setting metric perturbations to zero. See Ref.\cite{Salehian:2020bon}, in which higher-order corrections are considered, leading to an interesting effective-fluid description for the scalar field with nontrivial pressure and viscosity. 

The effect of the new terms beyond the SP equations can be explored by numerical simulations. To test the EFT in a simple (but still nontrivial) example, in the next section we study solitonic solutions to the above equations under the assumption of spherical symmetry, neglecting the expansion of universe and the additional fluid component. Before we do that, we end this section with a few remarks regarding our results. 

Note that from the bottom-up EFT point of view one can expect the appearance of all correction terms in Eq.~\eqref{eqftkg} but with unknown coefficients. However, it is not the case that all terms consistent with the symmetries appear: for example, terms like $\Phi_s\lambda|\psi_s|^2\psi_s$ or $H_s\nabla^2\psi_s$ did not appear. This can be thought as the consequence of the original theory with which we started, namely general relativity with a scalar field minimally coupled to gravity. One may expect new terms to appear if one considers a modified theory of gravity, which may also change the coefficients of terms already identified in Eqs.~(\ref{eqftkg})-(\ref{eftPsi}). It would be interesting to explore such a possibility, which is beyond the scope of this paper. Furthermore, note that a term proportional to $\lambda^2$ has appeared in Eq.~\eqref{eqftkg} with the same structure as the $\kappa$ term. Therefore, we see that a single term in the original theory gives rise a tower of terms in the low-energy EFT as a result of integrating out high-energy modes.  See Refs.~\cite{Namjoo:2017nia,Braaten:2018lmj} for more details.

Finally, we note that the  SP equations \eqref{eq:SP} and \eqref{eq:SPe} in an expanding universe possess a scaling symmetry \cite{PhysRevD.50.3650} in the following sense. If $\psi_s(t,\bf{x})$ and $\Phi_s(t,\bf{x})$ are a set of solutions, then $\eta\psi_s(\eta t,\sqrt{\eta}\bf{x})$ and $\eta\Phi_s(\eta t,\sqrt{\eta}\bf{x})$ are also solutions if we make the replacement $\lambda\rightarrow\lambda/\eta$, $\rho_s\rightarrow\rho_s\eta^2$ and $H_s\rightarrow \eta H_s$ for any constant $\eta$.  And the small parameters ($\epsilon_x,\epsilon_\lambda,\epsilon_g,\epsilon_\varphi$) get multiplied by $\eta$. In general, this particular scaling symmetry of the solutions is lost as we include corrections to the SP system.

\section{Approximate solitonic solutions}\label{sec:spherical}

Equations \eqref{eqftkg}-\eqref{eftPsi} of our EFT, which include relativistic corrections to the SP system, can be incorporated in many different contexts and the solutions will take different forms. In this section we study one of the simplest solutions: spherically symmetric, stationary solutions of the form
\eq{
	\label{psistat}
	\psi_s(t,r) = f(r) e^{i\mu t} ~,\,
}
with $\Phi_s=\Phi_s(r)$ in Eqs.~\eqref{eqftkg} and \eqref{eftPhi} and $\Psi_s=\Psi_s(r)$ in Eq.~\eqref{eftPsi}. As we will see below, this form corresponds to approximate solitonic solutions. Under spherical symmetry, the vector and tensor modes vanish identically. Moreover, by definition $\psi_s$ is a slowly varying function of time, so we must have $\mu/m\sim\epsilon_t\ll1$. In this section we ignore FLRW expansion, setting $a(t) = 1$, and also ignore contributions from the background fluid.   

The specific form of the field in Eq.~\eqref{psistat} resembles the wavefunction of stationary states in quantum mechanics. Although strictly speaking the field $\psi_s$ is not a wavefunction, its time evolution is governed by Eq.~\eqref{eqftkg} which, at leading order, resembles the conventional Schr\"{o}dinger equation. In the quantum-mechanical context, stationary solutions correspond to states with vanishing probability current and hence no actual time evolution. Similarly, the ansatz of Eq.~\eqref{psistat} corresponds to a time-independent energy density.\footnote{One crucial difference between our system and conventional quantum mechanics is that our system is nonlinear (even without relativistic corrections), so that a superposition of solutions fails to be a solution. As a result, unlike in quantum mechanics, one cannot express the general time evolution in terms of a superposition of various stationary states.} 

By using the stationary, spherically symmetric form of $\psi_s$ in Eq.~(\ref{psistat}) and the time independence of $\Phi_s$ in Eqs.~\eqref{eqftkg} and \eqref{eftPhi}, we find
\eqa{
	\label{eftkgsss}
	\begin{split}
	&\frac{\nabla^2f}{2m}-\Big(\Phi_s+\frac{\mu}{m}\Big)mf-\frac{\lambda f^3}{8m^2}\\
	&+\Big(3\Phi_s^2+\frac{4\mu}{m}\Phi_s+\frac{\mu^2}{2m^2}\Big)mf+\frac{\lambda f^3}{8m^2}\Big(2\Phi_s-\frac{\mu}{m}\Big)+\frac{3f^3}{16\Mpl^2}-\frac{\lambda^2f^5}{768m^5}=0+\order{\epsilon^4} 
	\end{split}
	}
and
\eqa{	
	\begin{split}
	\label{eftPhisss}
	&\nabla^2\Phi_s-\frac{mf^2}{2\Mpl^2} \\
	&-\frac{mf^2}{2\Mpl^2}\Big(-6\Phi_s-3\frac{\mu}{m}\Big)+\frac{\lambda f^4}{16m^2\Mpl^2}=0+\order{\epsilon^4}\,,
	\end{split}
 }                               
where for simplicity we have set $\kappa=0$. The above equations are the \emph{time-independent} nonrelativistic EFT system of equations; in both Eq.~(\ref{eftkgsss}) and (\ref{eftPhisss}), terms on the second line are smaller than those on the first by $\order{\epsilon}$ (while the first lines are already $\order{\epsilon^2}$ according to our power counting). Note the explicit appearance of the parameter $\mu$ in these equations. Since the stationary ansatz of Eq.~\eqref{psistat} removes all time derivatives from Eqs.~\eqref{eqftkg} and \eqref{eftPhi}, we have used the leading-order equations to remove all spatial derivatives in subleading terms, which yields multiple terms proportional to $\mu$ in the final result.\footnote{Some appropriate field redefinitions can remove $\mu$ completely from the equations, but change the asymptotic behavior of $\Phi$ to a nonzero constant. In this case $\Psi$ can no longer be replaced by $\Phi$ to the working order, so that both gravitational potentials must be solved simultaneously. Such a system is easier to solve numerically, and the plots in this section take advantage of this procedure. We discuss this further in Appendix \ref{app:EFT_details} and show the resulting simplified set of equations in Appendix \ref{app:NREFT2}, in which we also go one order higher in the EFT expansion.} Moreover, the $\nabla^4$ term in Eq.~\eqref{eqftkg}, which appears due to integrating out nonzero modes $X_\nu$ with $\nu \neq 0$, can also be removed by a similar manipulation. This removal of higher spatial derivatives makes the system more suitable for numerical calculations. We look for spatially localized, nodeless and regular solutions. That is, we demand that $f(r)$ and $\Phi_s(r)$ vanish fast enough at infinity; that $f'(0)=0$; and that the solutions are monotonic. Such solutions are expected to describe long-lived solitonic solutions. 

We wish to compare solutions of our EFT, Eqs.~\eqref{eftkgsss} and \eqref{eftPhisss}, with corresponding solutions of the SP equations, as well as  solutions of a fully relativistic theory. This will allow us to see whether the EFT equations provide an improvement over the SP equations. Before doing so, however, we must address two questions: (1) What solution of the Klein-Gordon equation corresponds to the solution of Eqs.~\eqref{eftkgsss} and \eqref{eftPhisss}? (2) What observable should we choose in order to compare the solutions?

To answer the first question we try to reconstruct the scalar field $\vp$, or equivalently $\psi$, from the knowledge of $\psi_s$ and nonzero modes $\psi_\nu$. From Eq.~\eqref{nonzero} for the stationary solution we have
\eq{
	\psi^{(1)}_\nu=\left[\frac{\mu}{2m}e^{-i\mu t}\delta_{\nu,2}+\frac{\lambda f^2}{48m^3}e^{3i\mu t}\delta_{\nu,-2}-\frac{\lambda f^2}{96m^3}e^{-3i\mu t}\delta_{\nu,4}\right]f\,,
}        
where we have used the leading-order Schr\"{o}dinger equation to simplify terms. Using $\psi=\psi_s+\sum_{\nu\ne0}\psi_\nu^{(1)}e^{i\nu m t}+\order{\epsilon^3}$ and Eq.~\eqref{redef1}, we see that the relativistic scalar field will take the form
\eq{
	\vp=\sqrt{\frac{2}{m}}\left[\vp_1\cos(\omega t)+\vp_3\cos(3\omega t)+\order{\epsilon^3}\right]\,,
} 
where the higher-order terms also include higher multiples of the frequency $\omega \equiv m-\mu$. We have defined time-independent coefficients $\vp_1=(1+\mu/2m)f$ and $\vp_3=\lambda f^3/96m^3$ at this working order. As a result, the specific form of Eq.~\eqref{psistat} implies a periodic solution for the scalar field with a period ${2\pi}/{\omega}$, and we must look for this type of solution in the relativistic theory; keeping in mind that the true relativistic solutions also include radiating modes (leading to deviations from periodicity) \cite{Fodor:2008du,Fodor:2019ftc,Zhang:2020bec}, which are not captured here. Such field configurations correspond to approximate solitonic solutions.

Solutions in which the field configuration is spatially localized, coherently oscillating in the core, and the configuration is exceptionally long lived, are well known and are called oscillons \cite{Bogolyubsky:1976yu,Gleiser:1993pt,Copeland:1995fq,Kasuya:2002zs,Amin:2010jq,Amin:2013ika,Fodor:2019ftc}, axion stars, scalar stars \cite{Seidel:1991zh,Visinelli:2017ooc,Chavanis:2017loo, Eby:2019ntd} depending on the context. They are approximate, time-dependent solitons of the relativistic theory. Such solitons are relevant in many cosmological contexts, both in the early and contemporary universe (for example, see \cite{Copeland:1995fq,  Amin:2010xe,Amin:2011hj,Gleiser:2011xj, Grandclement:2011wz, Lozanov:2017hjm,Hong:2017ooe,Ikeda:2017qev,Levkov:2018kau, Bond:2015zfa,Antusch:2017flz, Arvanitaki:2019rax, Brax:2020oye,Kawasaki:2020jnw,Amin:2020vja}). They owe their localization to gravitational interactions\cite{Seidel:1991zh, Seidel:1993zk, UrenaLopez:2001tw, Alcubierre:2003sx}, or self-interactions \cite{Amin:2019ums, Zhang:2020bec, Zhang:2020ntm} or a combination of both.

As for the second question, one important and reliable observable we have for solitonic solutions is their mass. We use the ADM definition of mass \cite{Arnowitt:1962hi} (see also Ref.~\cite{Gourgoulhon:2007ue}), which is the Schwarzschild mass for an observer at infinity,
\eq{
	\label{mass}
	M\equiv\lim_{r\to\infty}\frac{r^2\p_r\Psi}{G_N}=-\int_{0}^{\infty}\dd{r}4\pi r^2\left[T^0{}_0+3\Mpl^2e^{-2\Phi}\dot{\Psi}^2\right]e^{-5\Psi/2}\,,
}   
where $G_N$ is Newton's gravitational constant, related to the reduced Planck mass by $G_N = 1/ (8 \pi \Mpl^2)$. In the second equality we have used the Einstein field equations and the expression for the $G^0{}_0$ component of the Einstein tensor. Since the ADM mass is time independent, in the language of the mode expansion of Eq.~\eqref{eq:modexp}, it only depends on the slow mode: $M=\ev{M}=M_s$. As a result, it is also possible to compute $M$ with the help of EFT variables $f$ and $\Phi_s$ as
\eq{
	\label{masss}
	M_s=\int_{0}^{\infty}\dd{r}4\pi r^2\left[m f^2\Big(1-\frac{7\Phi_s}{2}\Big)+\frac{(\nabla f)^2}{2m}+\frac{\lambda f^4}{16m^2}\right]~,
} 
to working order. This means that, although the KGE and SP (plus suitable corrections) systems are two different theories, we can compare their solutions by demanding that they both yield the same solitonic mass. 

For a given mass, the radius is a good measure with which to compare SP (plus corrections) and KGE solutions, since it involves comparing the density of the two solutions and roughly shows how fast the fields decay with distance from the origin. The radius of the soliton can be defined as the distance enclosing most of the mass (the integrand of Eq.~\eqref{mass}). However, since the integrand of Eq.~\eqref{mass} depends on time, this definition results in an oscillating radius for the soliton. Whereas in most applications such oscillations may not be an issue, here we must treat the oscillations carefully. To avoid any confusion, we \emph{define} the radius $R_{95}$ as the distance enclosing 95 percent of the mass after taking the time average of the mass density:\footnote{If we first find the 95 percent radius and then take its time average we get a result which is the same as above at leading order but has higher order corrections as well.}
\eq{0.95 M=\int_0^{R_{95}}\langle [\hdots ]\rangle 4\pi r^2 dr \, ,}
where the term in square brackets is given by the corresponding term in the integrand of Eq.~(\ref{masss}). 

In practice, it is easier to fix the value of the field at the origin, rather than the mass, and find the corresponding solution in both theories. The mass and radius of the solution can then be obtained from Eq.~\eqref{masss}. This can be done for our EFT system at lowest order corresponding to the usual SP  equations with no corrections, as well as incorporating corrections at both ${\cal O} (\epsilon)$ and ${\cal O} (\epsilon^2)$ beyond SP. That is, we consider SP, SP$\times[1+\mathcal{O}(\epsilon)]$ and SP$\times [1+\mathcal{O}(\epsilon)+\mathcal{O}(\epsilon^2)\,]$ equations respectively. The latter theory has been made explicit in Appendix~\ref{app:NREFT2}.

Similarly, results for the KGE equations are obtained by following Ref.~\cite{Alcubierre:2003sx} and expanding the fields and gravitational potentials in terms of Fourier cosine series up to the frequency $4\omega$. We note that it is much simpler to do the shooting for solutions in the KGE system in spherical coordinates, whereas results in the effective theory are given in isotropic coordinates. The procedure for relating the two coordinate systems is discussed in Appendix \ref{app:coordinate}. We emphasize that, instead of comparing theories at some fixed central field amplitudes, we compare them using the mass-radius curve.

\fg{
	\centering
	\includegraphics[width=0.6\textwidth]{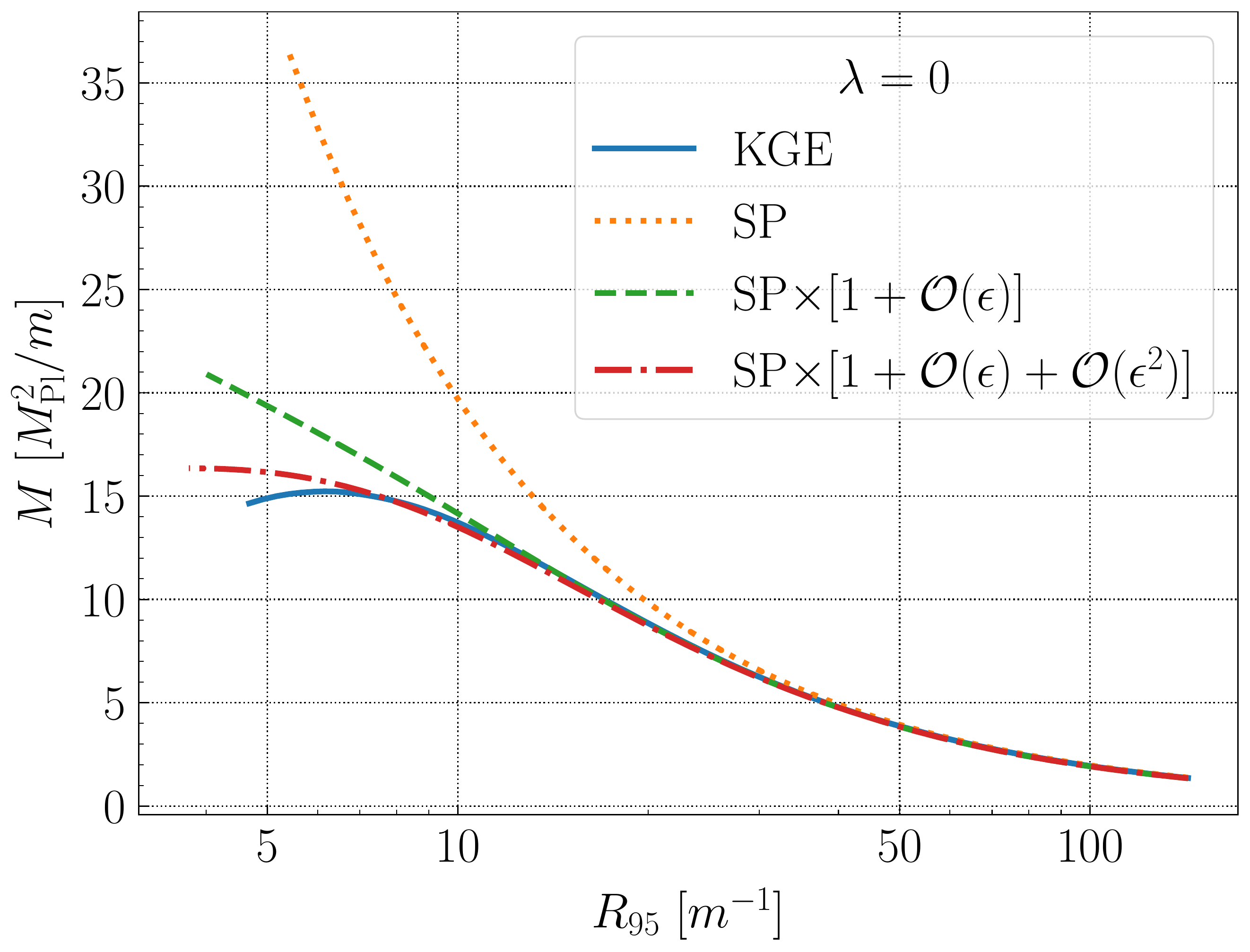} 
	\caption{A comparison of the mass-radius relation for the free theory ($\lambda = 0$) obtained from the Klein-Gordon-Einstein (KGE) equations, the Schr\"{o}dinger-Poisson (SP) equations, and our nonrelativistic effective field theory that includes $\mathcal O(\epsilon)$ and $\mathcal O(\epsilon^2)$ corrections beyond the SP equations. When the system becomes mildly relativistic, $R_{95}\sim10\,m^{-1}$, the SP equations show increasing disagreement with the fully relativistic results obtained from the KGE equations. On the other hand, our effective equations with just the leading relativistic corrections improve the results significantly. Furthermore, the $\mathcal O(\epsilon^2)$ corrections also capture the qualitative behavior when the system becomes highly relativistic.}
	\label{fig:free}
}

\fg{
	\centering
	\includegraphics[width=0.49\textwidth]{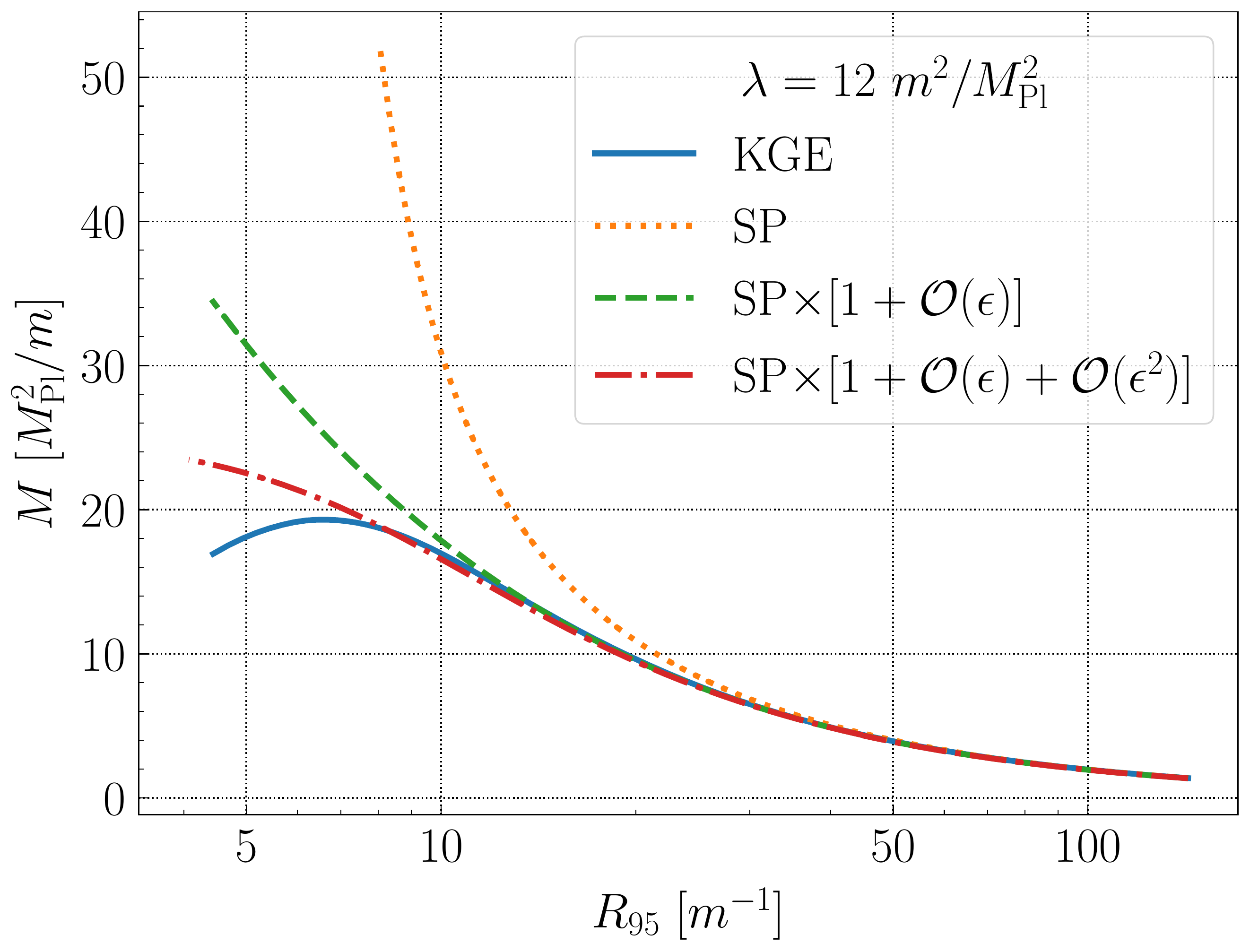} 
	\hspace{1mm}
	\includegraphics[width=0.49\textwidth]{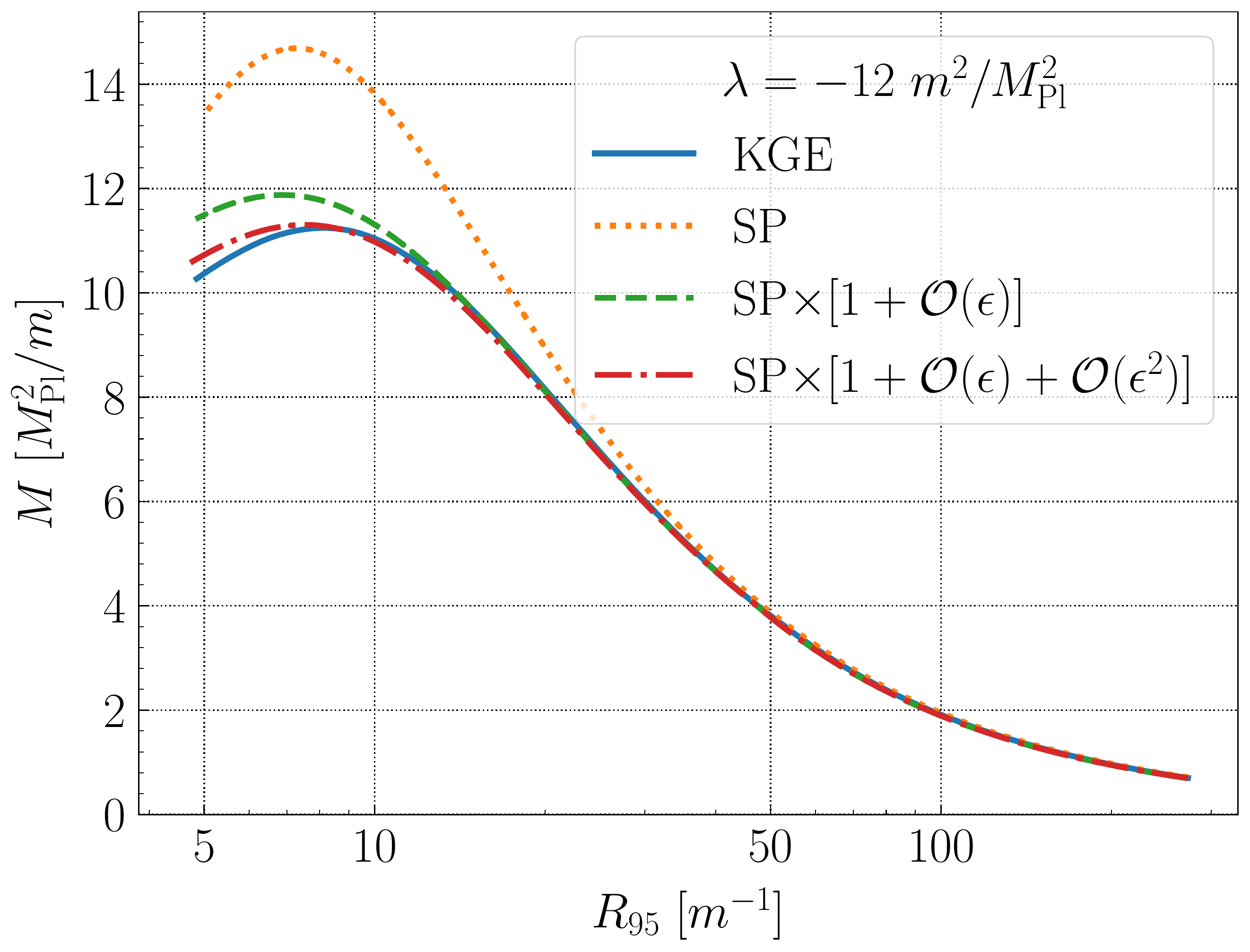}
	\caption{A comparison of the mass-radius relation for theories with repulsive (left) and attractive (right) self-interactions. The results for much larger $|\lambda\Mpl^2/m^2|$ (as would be the case for QCD and ultra-light axions) can also be obtained within our EFT. In this case the deviations from the SP system with attractive self-interactions still appear at $mR_{95}\lesssim 50$. For the repulsive case, the deviation from the SP system becomes significant at larger and larger $mR_{95}$ as $|\lambda\Mpl^2/m^2|$ increases.}
	\label{fig:int}
}

In Fig.~\ref{fig:free} we compare the mass-radius relationship for solitons in the free theory ($\lambda = 0$) obtained with the KGE and with SP equations (with and without corrections). In the figure, ``SP" refers to results from the lowest-order SP equations, neglecting all corrections, as given by the first lines of Eqs.~\eqref{eftkgsss} and \eqref{eftPhisss}. The EFT including $\order{\epsilon}$ corrections to the SP equations is given by Eqs.~\eqref{eftkgsss} and \eqref{eftPhisss} (first and second lines), and the $\order{\epsilon^2}$ corrections are given in Appendix \ref{app:NREFT2}. Compared with the SP equations, our effective equations with just the $\order{\epsilon}$ corrections improve the mass-radius relation significantly in the mildly relativistic regime, for $R_{95}\simeq 10 ~m^{-1}$. Note that at this radius, the mass calculated using the SP equations differs from that obtained from the relativistic KGE calculation by $>50\%$, whereas including $\order{\epsilon}$ corrections in our EFT leads to a discrepancy of $<10\%$ from the relativistic KGE solutions. We can improve the results further by including $\order{\epsilon^2}$ corrections, which match the relativistic KGE calculation to within $\sim 1\%$ around $R_{95} \simeq 10~m^{-1}$.

We have also confirmed the improvement in the mass-radius relationship obtained from our equations compared to the SP system in theories with repulsive ($\lambda>0$) and attractive ($\lambda<0$) self-interactions in Fig.~\ref{fig:int}. To check the validity of our effective theory, we plot various small parameters $\epsilon$ for the theory with repulsive self-interactions in Fig.~\ref{fig:epsilon}. The left panel shows the profile of $\epsilon$ in terms of the radius, while the right panel shows the maximum value of $\epsilon$ in terms of the 95\% radius. Our perturbative scheme fails when $\epsilon \sim 1$. Also note that the value of $\epsilon_x$ is a measure of particle momentum, and we see that particles are indeed mildly relativistic when $R_{95}\simeq 10 ~m^{-1}$.

We note that our reasons for choosing $\lambda=\pm 12 m^2/\Mpl^2$  in Fig.~\ref{fig:int} are that (i) by this choice all small parameters become the same order of magnitude in the mildly relativistic regime and (ii) it can make the comparison with the $\lambda=0$ easier. With more canonical parameter choices, the value of $|\lambda\Mpl^2/m^2|$ can be very large, for example, for QCD axions $\lambda\Mpl^2/m^2\sim -\Mpl^2/f_a^2$ where $f_a\sim 10^{11} \mathrm{GeV}$.  A natural question is at what $mR_{95}$ does the mass-radius relation of SP equations start to deviate significantly from that of KGE equations when $|\lambda\Mpl^2/m^2|\gg 10$? We have confirmed that few percent level differences from the SP system always start appearing at $mR_{95}\lesssim 50$ as long as $\lambda\Mpl^2/m^2\ll -10$.\footnote{In practice, as a proxy for the detailed mass radius curve, we simply construct $mR_{95}$ vs. $|\lambda|$ for $\epsilon_\lambda=0.1$, and see that this $mR_{95}$ initially grows slowly with $|\lambda|$ and then approaches a constant at sufficiently large $|\lambda|$.}

For the repulsive case, and with $\lambda\Mpl^2/m^2\gg 10$, there is a minimum radius for the soliton (in the SP system). Around the minimum, the soliton is formed by a balance between gravity and self-interactions which implies that $mR_\mathrm{min}\propto \sqrt{\lambda\Mpl^2/m^2}$. The deviation between results from SP and KGE equations is large at $mR_\mathrm{min}$. As a result, relativistic corrections become important at larger $mR_{95}$ as we increase the value of $\lambda\Mpl^2/m^2$ -- which shifts the minimum (and therefore the whole curve) to the right in the mass-radius plot. The difference between attractive and repulsive cases is because the former has a balance between the gradient term and attractive self-interactions. For the repulsive case, also see Ref. \cite{Croon:2018ybs}.

\fg{
	\centering
	\includegraphics[width=0.49\textwidth]{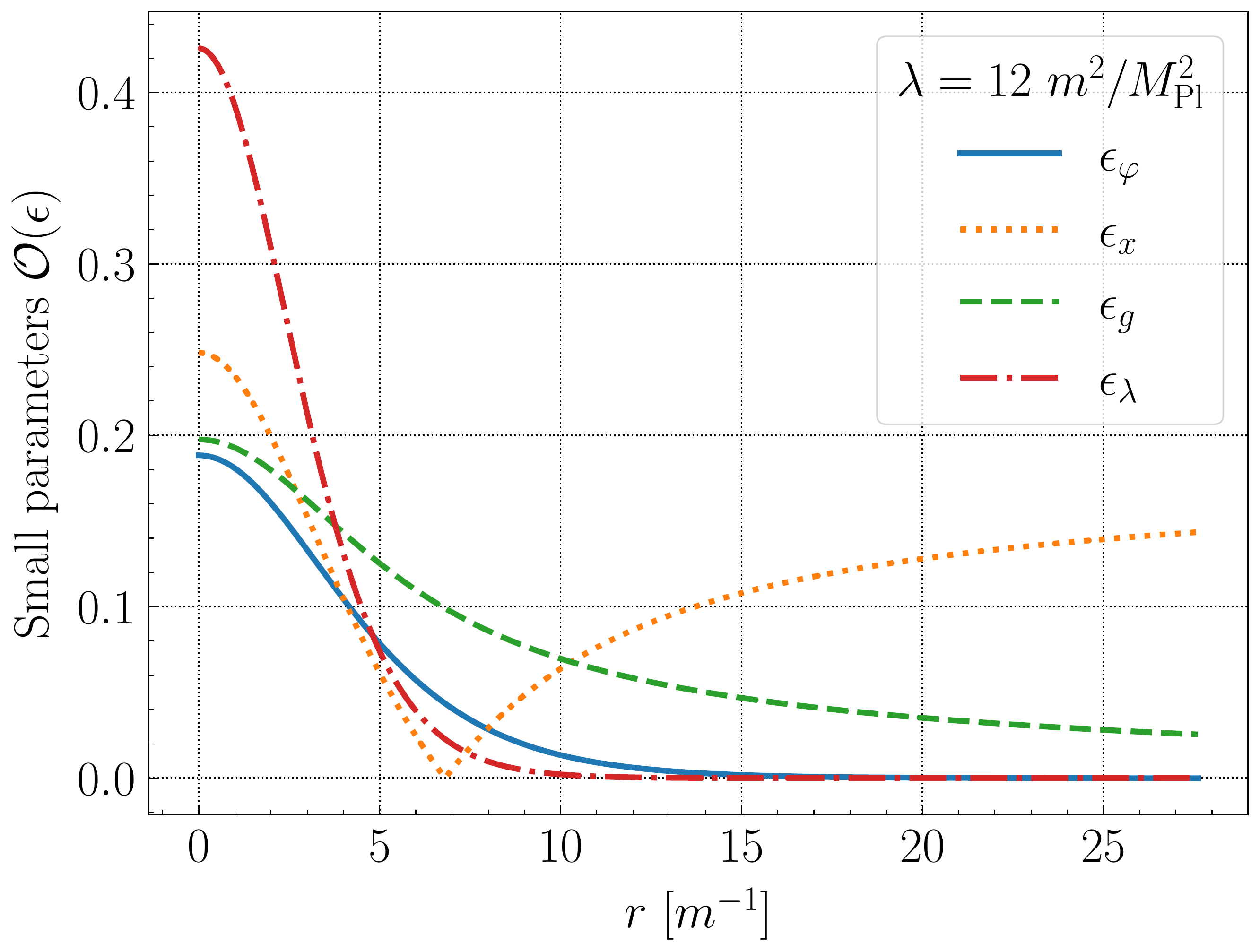} 
	\hspace{1mm}
	\includegraphics[width=0.49\textwidth]{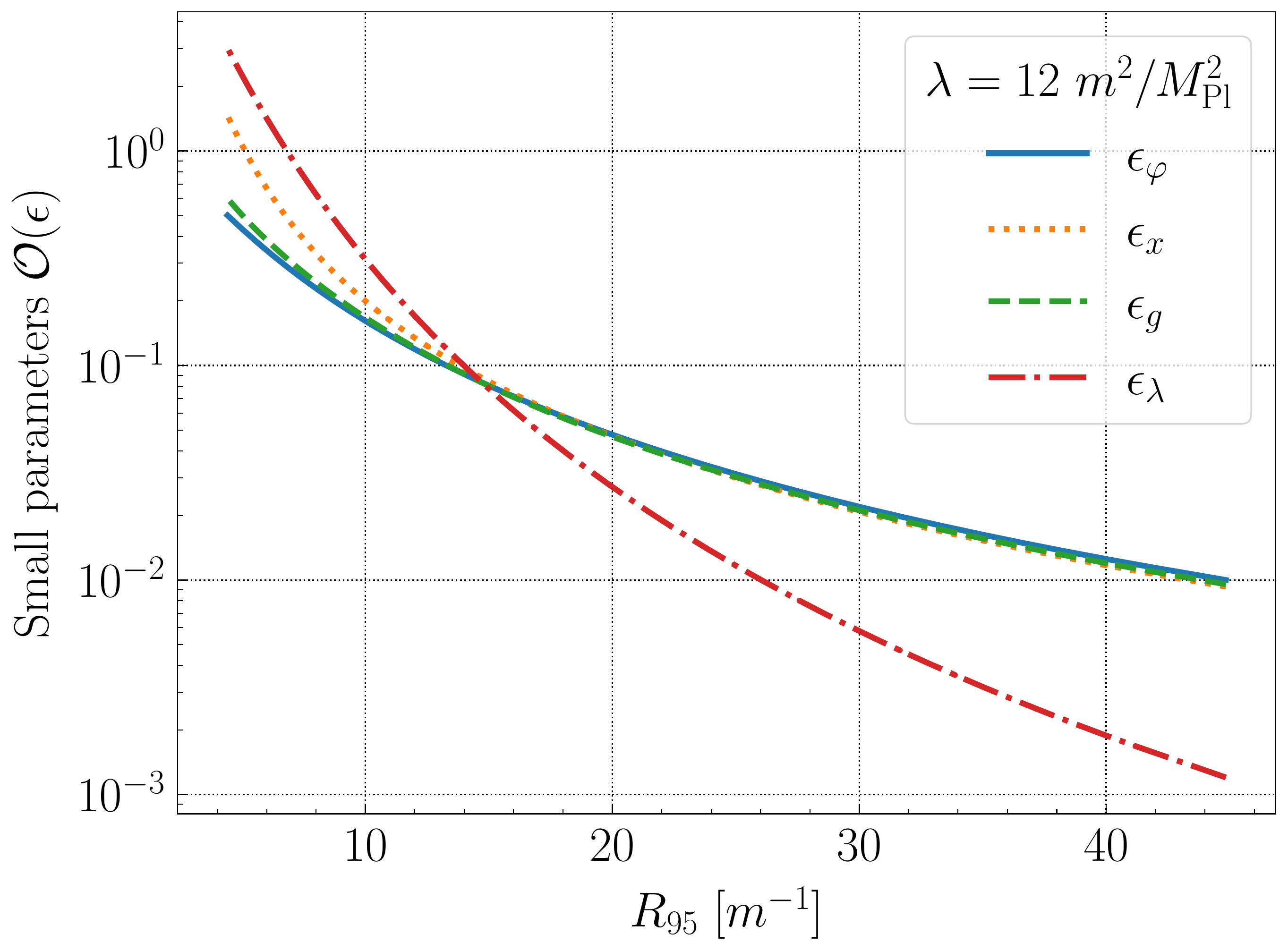}
	\caption{Small parameters in the effective field theory for the soliton solutions. In the left panel, we plot the spatial dependence of the small parameters $\epsilon$ for a soliton with $\tilde f(r=0)=0.19$ or $R_{95}=9.08 ~m^{-1}$, where $\tilde{f} (r)$ is defined in Eq.~(\ref{redef}). In the right panel, we plot the maximum value of $\epsilon$ as a function of the 95\% radius. The data is obtained based on our effective equations with $\mathcal O(\epsilon^2)$ corrections. By definition our small parameters $\epsilon$ are positive definite, see Sec.~\ref{sec:FLRW} and \ref{sec:power-counting}.}
	\label{fig:epsilon}
} 

\section{Summary and conclusion}\label{sec:summary}
In this article we have provided a nonrelativistic effective field theory (EFT) for scalar dark matter in a cosmological context. We obtained the EFT by systematically integrating out 
rapidly oscillating modes from our system. We have treated matter nonperturbatively while keeping metric perturbations to the desired order in the EFT (including nonlinear metric perturbations). Our approach recovers the familiar Schr\"{o}dinger-Poisson system at the lowest order in the EFT and provides explicit expressions for the corrections due to scalar field self-interactions and general-relativistic corrections, as identified in Eqs.~(\ref{eqftkg}) - (\ref{eftPsi}).

Our framework allows for a systematic assessment of the relative importance of the various corrections to the Schr\"{o}dinger-Poisson (SP) system in the mildly relativistic regime. For example, we showed that for leading-order corrections to the SP system, it is sufficient to include scalar modes at sub-leading order and vector modes at leading order, while neglecting tensor modes. As another example, if self-interactions are sufficiently strong it might be possible to ignore gravitational corrections \cite{Amin:2019ums}. Furthermore, a straightforward consequence of our results is that we see the loss of a particular scaling symmetry present in the SP system when we include leading-order corrections to the SP system (even with the quartic self-coupling set to zero). Our expressions also reveal that not all possible correction terms that might be expected from a bottom-up effective field theory consideration are present in general-relativistic systems. This result might be relevant for constraining departures from general relativity.

On the one hand our work provides a robust justification for using the Schr\"{o}dinger-Poisson equations in many contexts relevant for structure formation. On the other hand, it provides guidance about how to improve upon the Schr\"{o}dinger-Poisson system when relativistic corrections begin to become important without resorting to a fully relativistic analysis. Even when incorporating relativistic corrections, our EFT approach does not require tracking  (or resolving) short time-scale dynamics of the system. The equations for the slowly varying, nonrelativistic modes acquire modifications upon integrating out the rapidly oscillating modes.  

Implications of the sub-leading terms in the EFT can become important in the mildly relativistic regime. This could arise because one or all of the small parameters (defined in Section \ref{sec:power-counting}) are not sufficiently small. For example,  we used our EFT equations to derive the mass-radius relationship of spherically symmetric dense solitons. We were able to obtain a significant improvement in the accuracy of this curve by using our EFT to go beyond the Schr\"{o}dinger-Poisson system. Our EFT with corrections is simpler to use than the fully relativistic treatment in the mildly relativistic regime. However, as a limitation, our EFT cannot capture the decay of solitons by number-changing processes \cite{Hertzberg:2010yz,Mukaida:2016hwd,Eby:2018ufi,Zhang:2020bec}.

Whereas we only studied the stationary soliton in this paper, the time-dependent problem of mildly relativistic collapse is the next logical step to be studied with our EFT (see Ref.~\cite{Guzman:2003kt} for non-relativistic collapse). This study is possible within our EFT primarily because there is a hierarchy between the scale associated with nonrelativistic gravitational collapse, given by the Jeans scale ($\lambda_J$), and the Compton length-scale ($\lambda_C$) at which relativistic corrections become important: $\lambda_J\gg \lambda_C$. Equivalently, one can see that many scalar-field oscillations with period similar to the Compton time ($\tau_C$) fit within the Jeans time ($\tau_J$) associated with matter collapse.\footnote{ The Jeans length scales as $\lambda_C\sim\sqrt{\varrho}/\Mpl$, where $\varrho\approx m|\psi|^2$ is the matter density, whereas the Compton scale is given by $\lambda_C\sim 1/m$. Similarly, the collapse time scales as $\tau_J\sim \sqrt{\varrho}/\Mpl$ whereas the oscillation period is given by $\tau_C\sim 1/m$.} In certain regimes, the collapse can be protected from becoming too relativistic by gradients and anisotropies in the field. 

Beyond spherical collapse, we can also employ our EFT to study transient vortices in fuzzy dark matter \cite{Hui:2020hbq}. Our results might be useful for studying linear and nonlinear structure formation in axion-like fields in the early universe, and for exploring the gravitational and self-interaction driven clustering dynamics of a weakly coupled inflaton field after the end of inflation \cite{Amin:2019ums,Musoke:2019ima} 
(or, more generally, moduli fields \cite{Kane:2015jia}). Whereas SP systems are prominently used in late-universe applications, most studies of the early universe use a nonlinear Klein-Gordon equation in an expanding universe, with occasional inclusion of gravitational perturbations \cite{Lozanov:2019ylm,Giblin:2019nuv}. Our work can also be seen as a link connecting the nonrelativistic SP system with the partially relativistic system used in the early-universe context. Our EFT incorporates nonlinear aspects of gravity, and can thus provide a stepping stone towards exploring systems (such as scalar field collapse to black holes \cite{Widdicombe:2018oeo, Nazari:2020fmk, Helfer:2018vtq}), for which it may be necessary to deploy the full power of general relativity. 

On the more formal side, it would be interesting to consider an effective fluid description of our corrections to the SP system using the Madelung-like transforms \cite{1927ZPhy...40..322M,Suarez:2015fga, Cookmeyer:2019rna,Salehian:2020bon}. Finally, repeating our work for more general matter content (higher-spin fields, such as vector dark matter \cite{Graham:2015rva, Adshead:2021kvl}), or different relativistic theories of gravity \cite{Ferreira:2019xrr}, could change the nature of the corrections to the SP system (or even the SP system itself). This could further allow for constraining the nature of the underlying fields that constitute dark matter.

\subsection*{Acknowledgments}
We would like to thank Mudit Jain, Siyang Ling, Andrew Long and Zong-Gang Mou (Rice University) and JiJi Fan (Brown University) for helpful discussions. Portions of this work were conducted in MIT's Center for Theoretical Physics and supported in part by the U.S.~Department of Energy under Contract No.~DE-SC0012567. MA is supported by a NASA ATP theory grant NASA-ATP Grant No. 80NSSC20K0518.
  
\appendix

\section{Equations for a perturbed FLRW universe}\label{equations}
In this Appendix we list the set of equations for a scalar field and an additional homogeneous and isotropic perfect fluid in a perturbed FLRW universe. Compared to what appeared in Sec.~\ref{sec:equations}, we keep all scalar modes nonperturbatively while we only consider linear corrections to the equations due to the vector mode. We derive these equations with two limits in mind, for which such a framework is self-consistent: In Sec.~\ref{eftnr} we proceed one order beyond the SP system, for which our equations below reduce to Eqs.~\eqref{eq:psidotD3}-\eqref{eqsigma-simp}, and in Sec.~\ref{sec:spherical} and Appendix \ref{app:NREFT2} we work under the assumption of spherical symmetry (in which case the vector mode would not be excited). For the same reason, we also neglect tensor modes, as a result of estimates discussed in Sec.~\ref{sec:power-counting}. With these considerations in mind, from the Einstein field equations we obtain
\eq{
	\label{eqPhi}
	\spl{
		&\frac{\nabla^2\Phi}{a^2}+\frac{\nabla(\Phi-\Psi) \cdot \nabla\Phi}{a^2}+3e^{-2(\Phi+\Psi)}\left(H\dot{\Phi}+2H\dot{\Psi}-\dot{\Phi}\dot{\Psi}-\dot{\Psi}^2+\ddot{\Psi}-\frac{\ddot{a}}{a}\right)+\order{\epsilon^{5/2}\sigma_j,\epsilon^{2}h_{ij}}\\
		&=\frac{e^{-2\Psi}}{2\Mpl^2}\Big[(\rho+3p)+\mathcal{S}_\Phi\Big]\,,}}
\eq{
	\label{eqPsi}
	\frac{\nabla^2\Psi}{a^2}-\frac{(\nabla\Psi)^2}{2a^2}+\frac{3}{2}e^{-2(\Phi+\Psi)}\left(H^2+\dot{\Psi}^2-2H\dot{\Psi}\right)+\order{\epsilon^{5/2}\sigma_j,\epsilon^{2}h_{ij}}=\frac{e^{-2\Psi}}{2\Mpl^2}\Big[\rho+\mathcal{S}_\Psi\Big]\,,
}
\eq{
	\label{G0j}
	\spl{
		\p_j\dot{\Psi}+H\p_j\Phi-\dot{\Psi}\p_j\Phi-&\frac{\nabla^2\sigma_j}{4a}\\&=-\frac{i}{4\Mpl^2} \left(e^{-imt}\psi-e^{imt}\psi^* \right) \p_j \left(e^{-imt}\psi+e^{imt}\psi^* \right) +\order{\epsilon^2\sigma_j,\epsilon^{5/2}h_{ij}}\,,
}}
\eq{
	\label{eqsigma}
	\frac{\nabla^4\sigma_j}{a}=\frac{2i}{\Mpl^2}\Big[\left( \nabla^2\psi+(\nabla\psi \cdot \nabla) \right) \p_j\psi^* - \left(\nabla^2\psi^*+(\nabla\psi^* \cdot\nabla) \right) \p_j\psi\Big]+\order{\epsilon^3\sigma_j,\epsilon^{7/2}h_{ij}}\,,
}
where we have defined
\eq{
	\label{Sphi}
	\spl{
		\mathcal{S}_\Phi=&-\frac{m}{2} \left( e^{-imt}\psi+e^{imt}\psi^* \right)^2 - m\,e^{-2\Phi} \left( e^{-imt}\psi-e^{imt}\psi^* \right)^2 - \frac{\lambda}{2\times4!m^2} \left( e^{-imt}\psi+e^{imt}\psi^* \right)^4\\
		&-\dfrac{\kappa}{4\times 6! \Lambda^2 m^3} \left( e^{-imt}\psi-e^{imt}\psi^* \right)^6+\order{m\psi^2\epsilon^{1/2}\sigma_j}\, ,
}}
\eq{
	\label{Spsi}
	\spl{
		\mathcal{S}_\Psi=&\frac{m}{4} \left( e^{-imt}\psi+e^{imt}\psi^* \right)^2 - \frac{m}{4}\,e^{-2\Phi} \left( e^{-imt}\psi-e^{imt}\psi^* \right)^2 +\frac{e^{2\Psi}}{4ma^2} \left( e^{-imt}\nabla\psi + e^{imt}\nabla\psi^* \right)^2 \\
		&+\frac{\lambda}{4\times4!m^2} \left( e^{-imt}\psi+e^{imt}\psi^* \right)^4 + \dfrac{\kappa}{8\times 6! \Lambda^2 m^3} \left( e^{-imt}\psi-e^{imt}\psi^* \right)^6+\order{m\psi^2\epsilon\,h_{il}}\,.
}}
Note that Eq.~\eqref{eqsigma} is obtained by applying the operator  $(\p_i\p^j-\delta_i{}^j\nabla^2)$ to Eq.~\eqref{G0j}, so Eq.~\eqref{eqsigma} is not an independent equation. 

The equation for $\psi$ can be obtained from Eq.~\eqref{eq:psidotD}, which, after some simplification, results in
\eq{
	\label{eq:psidotD2}
	i\dot{\psi}+\tilde{\mathcal{D}}\psi+e^{2imt}\tilde{\mathcal{D}}^*\psi^*=e^{imt}e^{2\Phi}\mathcal{J}\,,
}
with
\eqa{
	\label{D}
	\tilde{\mathcal{D}}&= \frac{m}{2} \left( 1-e^{2\Phi} \right) + \frac{e^{2(\Phi+\Psi)}}{2ma^2}\left( \nabla^2+\nabla(\Phi-\Psi) \cdot \nabla\right)+\frac{3i}{2}\left[H-\frac{\dot{\Phi}}{3}-\dot{\Psi}\right]-\frac{i}{a}\vec{\sigma}.\nabla+\order{\epsilon^{3/2}\sigma_j,\epsilon h_{ij}}\,,\\
	\label{J}
	\mathcal{J}&=\frac{\lambda}{4\times3!m^2} \left( e^{-imt}\psi+e^{imt}\psi^* \right)^3 + \frac{\kappa}{8\times5!\Lambda^2m^3} \left( e^{-imt}\psi+e^{imt}\psi^* \right)^5\,.
}
For the homogeneous and isotropic background we have 
\ba 
\label{eq:bg_full}
i\dot{\bar{\psi}}+\frac{3i}{2}H \left( \bar{\psi}-e^{2imt}\bar{\psi}^* \right) = e^{imt}\mathcal{\bar{J}}, \, \quad 
3\Mpl^2H^2=\rho+\bar{\mathcal{S}}_\Psi, \, \quad 
\frac{\ddot{a}}{a}=-\frac{1}{6\Mpl^2}\left[(\rho+3p)+\bar{\mathcal{S}}_\Phi\right].
\ea 
Based on the power counting discussed in Sec.~\ref{sec:power-counting}, one can check that the above relations reduce to Eqs.~\eqref{eq:psidotD3}-\eqref{bg} as far as the leading-order corrections to the SP system are concerned.

\section{ Details of the EFT derivation}
\label{app:EFT_details}
In this Appendix we provide some details for obtaining the effective equations in the nonrelativistic limit. We derive the leading-order corrections here while in Appendix \ref{app:NREFT2} we obtain next-to-leading-order corrections for the spherically symmetric solitonic solutions. 

\subsection{EFT for the background}
As a warm up, let us start with the equations for the spatially homogeneous background quantities which, as far as the leading-order corrections are concerned, are given by Eq.~\eqref{bg}. Note that Eq.~\eqref{bg} is slightly simplified compared to Eq.~\eqref{eq:bg_full} in that the self-interaction $\kappa$ term is neglected except in the Schr\"odinger equation.

To obtain the effective equations for the slowly varying part of the variables, we start by applying the mode expansion described in Sec.~\ref{eftnr} to the background equations and obtain equations for each mode $\nu$ as follows:
\eq{
	\spl{
		&i\big(\dot{\bar{\psi}}_\nu+i\nu m\bar{\psi}_\nu\big)+\frac{3i}{2}H_\alpha(\bar{\psi}_{\nu-\alpha}-\bar{\psi}^*_{2+\alpha-\nu})\\
		&=\frac{1}{24m^2}\Bigg[\lambda\Big((\bar{\psi}^3)_{\nu+2}+(\bar{\psi}^3)^*_{4-\nu}+3\bar{\psi}^*_{-\alpha}(\bar{\psi}^2)_{\nu-\alpha}+3\bar{\psi}_{\alpha}(\bar{\psi}^2)^*_{2+\alpha-\nu}\Big)+\frac{\kappa}{40m\Lambda^2}\Big((\bar{\psi}^5)_{\nu+4}+(\bar{\psi}^5)^*_{6-\nu}\\
		&+5\bar{\psi}^*_{-\alpha}(\bar{\psi}^4)_{\nu+2-\alpha}+5\bar{\psi}_{\alpha}(\bar{\psi}^4)^*_{4+\alpha-\nu}+10(\bar{\psi}^2)_{\alpha}(\bar{\psi}^3)^*_{2+\alpha-\nu}+10(\bar{\psi}^2)^*_{-\alpha}(\bar{\psi}^3)_{\nu-\alpha}\Big)\Bigg]+\dots\,,
	}
}
where summation over repeated indices is understood, and we have 
\eq{
	(\bar{\psi}^{n+1})_\nu=\sum_{\alpha_1,\alpha_2,\dots,\alpha_n}\bar{\psi}_{\alpha_1}\times\bar{\psi}_{\alpha_2}\times\dots\times\bar{\psi}_{\alpha_n}\times\bar{\psi}_{\nu-\alpha_1-\alpha_2-\dots-\alpha_n}\,,
}
and we recall that overbars denote spatially homogeneous quantities, $\bar{X} \rightarrow \bar{X} (t)$.
Similarly, for the other two equations we have
\eq{
	\spl{
		3\Mpl^2H_\alpha H_{\nu-\alpha}=\rho_\nu+m\bar{\psi}^*_{-\alpha}\bar{\psi}_{\nu-\alpha}+\frac{\lambda}{96m^2}\Big[&(\bar{\psi}^4)_{\nu+4}+(\bar{\psi}^4)^*_{4-\nu}+4\bar{\psi}^*_{-\alpha}(\bar{\psi}^3)_{\nu+3-\alpha}\\&+4\bar{\psi}_{\alpha}(\bar{\psi}^3)^*_{3+\alpha-\nu}+6(\bar{\psi}^2)_{\alpha}(\bar{\psi}^2)^*_{\alpha-\nu}\Big]+\dots\,,
}}
and 
\eq{
	\big(\dot{\rho}_\nu+i\nu m \rho_\nu)+3H_\alpha\big(\rho_{\nu-\alpha}+p_{\nu-\alpha}\big)=0\,.
}
We are interested in the system of equations for the slow modes ($\nu=0$). However, the slow modes are coupled to infinitely many nonzero modes (i.e., modes with $\nu \neq 0$). Fortunately, due to the nonrelativistic nature of the system there is a hierarchy among different quantities, which enables us to solve for the nonzero modes perturbatively. The formal solution for the nonzero modes ($\nu \neq 0$) is
\eq{
	\rho_\nu=\frac{i\dot{\rho}_\nu}{\nu m}+\frac{3i}{\nu m}H_\alpha\big(\rho_{\nu-\alpha}+p_{\nu-\alpha}\big)\,,
}
\eq{
	\spl{
		\bar{\psi}_\nu=&\frac{i\dot{\bar{\psi}}_\nu}{\nu m}+\frac{3i}{2\nu m}H_\alpha(\bar{\psi}_{\nu-\alpha}-\bar{\psi}^*_{2+\alpha-\nu})\\
		&-\frac{i}{24\nu m^3}\Bigg[\lambda\Big((\bar{\psi}^3)_{\nu+2}+(\bar{\psi}^3)^*_{4-\nu}+3\bar{\psi}^*_{-\alpha}(\bar{\psi}^2)_{\nu-\alpha}+3\bar{\psi}_{\alpha}(\bar{\psi}^2)^*_{2+\alpha-\nu}\Big)\\
		&+\frac{\kappa}{40m\Lambda^2}\Big((\bar{\psi}^5)_{\nu+4}+(\bar{\psi}^5)^*_{6-\nu}+5\bar{\psi}^*_{-\alpha}(\bar{\psi}^4)_{\nu+2-\alpha}+5\bar{\psi}_{\alpha}(\bar{\psi}^4)^*_{4+\alpha-\nu}\\
		&\hspace{50mm}+10(\bar{\psi}^2)_{\alpha}(\bar{\psi}^3)^*_{2+\alpha-\nu}+10(\bar{\psi}^2)^*_{-\alpha}(\bar{\psi}^3)_{\nu-\alpha}\Big)\Bigg]\,,
	}
}
and 
\eq{
	\spl{
		H_\nu=\frac{1}{6\Mpl^2H_s}\Bigg[\rho_\nu&+m\bar{\psi}^*_{-\alpha}\bar{\psi}_{\nu-\alpha}+\frac{\lambda}{96m^2}\Big[(\bar{\psi}^4)_{\nu+4}+(\bar{\psi}^4)^*_{4-\nu}+4\bar{\psi}^*_{-\alpha}(\bar{\psi}^3)_{\nu+3-\alpha}\\&\hspace{10mm}+4\bar{\psi}_{\alpha}(\bar{\psi}^3)^*_{3+\alpha-\nu}+6(\bar{\psi}^2)_{\alpha}(\bar{\psi}^2)^*_{\alpha-\nu}\Big]-3\Mpl^2\sum_{\alpha\neq0,\nu}H_\alpha H_{\nu-\alpha}\Bigg]\,.
}}
Note that in the formal solutions above we have used the fact that all modes, $\bar{\psi}_\nu$ and $\rho_\nu$, are slowly varying and their time derivatives are suppressed by at least one factor of $\epsilon$. In fact, all terms on the right-hand side are formally suppressed by at least one factor of $\epsilon$ compared to the slow mode of the variable appearing on the left-hand side.  

By using the power counting presented in Sec.~\ref{sec:power-counting} one can find solutions for the nonzero modes at leading order as follows:
\eqa{
	\label{psibnu}
	\bar{\psi}^{(1)}_\nu&=\left(-\frac{3iH_s}{4m}-\frac{\lambda|\bar{\psi}_s|^2}{16m^3}\right)\bar{\psi}_s^*\delta_{\nu,2}+\frac{\lambda\bar{\psi}_s^3}{48m^3}\delta_{\nu,-2}-\frac{\lambda\bar{\psi}_s^*{}^3}{96m^3}\delta_{\nu,4}\\
	\label{Hnu}
	H^{(1)}_\nu&=-\frac{i\bar{\psi}_s^*{}^2}{8\Mpl^2}\delta_{\nu,2}+\frac{i\bar{\psi}_s^2}{8\Mpl^2}\delta_{\nu,-2} \, , \qquad \rho^{(1)}_\nu=0\,.
}  
Note that at leading order, there are only a finite number of nonzero modes that remain nonvanishing. Next we use the above results to remove the nonzero modes that appear in the equations for the slow modes, $\bar{\psi}_s=\ev{\bar{\psi}}$, $H_s=\ev{H}$, and $\rho_s=\ev{\rho}$, which results in
\eqa{
	&i\dot{\bar{\psi}}_s+\frac{3i}{2}H_s\bar{\psi}_s-\frac{\lambda|\bar{\psi}_s|^2}{8m^2}\bar{\psi}_s+\textcolor{black}{\frac{(6\rho_s+9m|\bar{\psi}_s|^2)}{16m\Mpl^2}\bar{\psi}_s}+\textcolor{black}{\left(\frac{17\lambda^2}{8m^2}-\frac{\kappa}{\Lambda^2}\right)\frac{|\bar{\psi}_s|^4}{96m^3}\bar{\psi}_s}+\dots=0\\
	&3\Mpl^2H_s^2=\rho_s+m|\bar{\psi}_s|^2+\textcolor{black}{\frac{\lambda|\bar{\psi}_s|^4}{16m^2}}+\dots \, , \qquad \dot{\rho}_s+3H_s(\rho_s+p_s)=0\,,
}
where dots stand for higher-order terms. It is instructive to have a comparison with the naive equations in which one neglects all the oscillating terms in Eq.~\eqref{eq:bg_full} and obtains
\eqa{
	\nonumber
	i\dot{\bar{\psi}}+\frac{3i}{2}H\bar{\psi}-\frac{\lambda|\bar{\psi}|^2}{8m^2}\bar{\psi}-\frac{\kappa|\bar{\psi}_s|^4}{96\Lambda^2m^3}\bar{\psi}_s\simeq 0\qquad\nonumber3\Mpl^2H^2\simeq \rho+m|\bar{\psi}|^2+\textcolor{black}{\frac{\lambda|\bar{\psi}|^4}{16m^2}}\qquad\nonumber\dot{\rho}+3H(\rho+p)\simeq 0\,.
}
In comparison, there are several terms that appear in the systematic EFT expansion of the Schr\"odinger equation. The effect of these terms becomes more and more important as the system becomes more and more relativistic, that is, when small parameters, denoted by $\epsilon$, grow. The two other equations do not get any explicit corrections at this order. This does not mean that the behavior of $H_s$ and $\rho_s$ are not affected at this order. Rather, due to the coupling to $\bar{\psi}_s$ they deviate from the naive solution. Furthermore, by considering next-to-leading-order corrections, one does find new terms appearing in the equations for the Hubble parameter and for the fluid (see Ref.~\cite{Salehian:2020bon}).

One can obtain an effective equation for the slow mode of the scale factor by a similar procedure. The easiest way to do this is to start from the definition of the Hubble parameter and employ the mode decomposition, which results in
\eq{
	\dot{a}=aH\qquad\implies \qquad \dot{a}_\nu+i\nu m a_\nu=H_\alpha a_{\nu-\alpha}\,.
}
It is then easy to see that $a^{(1)}_\nu=0$, hence the equation for the slow-mode at the working order is $\dot{a}_s=a_sH_s+\dots$. 

\subsection{Inhomogeneities} 
We now turn to the EFT for inhomogeneities. The procedure is similar to what we have done for the background theory. In contrast to what is done in Sec.~\ref{eftnr}, here we keep $\Phi$ and $\Psi$ distinguished for future convenience of field redefinitions. This allows us to remove one parameter from the theory governing spherically symmetric solitons. This can be done by a redefinition of variables, as shall be discussed shortly. 

We start by a formal mode decomposition of the equations for inhomogeneities, Eqs.~\eqref{eqPhi}-\eqref{eqsigma}, which results in
\eqa{
&i\big(\dot{\psi}_\nu+i\nu m\psi_\nu\big)+\tilde{\mathcal{D}}_\alpha\psi_{\nu-\alpha}+\tilde{\mathcal{D}}^*_{-\alpha}\psi^*_{2+\alpha-\nu}=\big(e^{2\Phi}\big)_\alpha\big(e^{imt}\mathcal{J}\big)_{\nu-\alpha}\\
&\left(\frac{1}{a^2}\right)_\alpha\big(\nabla^2\Phi_{\nu-\alpha}+\nabla(\Phi_\beta-\Psi_\beta) \cdot \nabla\Phi_{\nu-\alpha-\beta}\big)\\\nonumber&\hspace{5mm}-3\big(e^{-2(\Phi+\Psi)}\big)_{\nu-\alpha}\left(\big(\chi^2\big)_\alpha+\dot{\chi}_\alpha-i\alpha m\chi_\alpha-\big(\dot{\Phi}\chi\big)_\alpha\right)=\frac{1}{2\Mpl^2}\big(e^{-2\Psi}\big)_{\nu-\alpha}\Big[(\rho_\alpha+3p_\alpha)+\big(\mathcal{S}_\Phi\big)_\alpha\Big]\\
&\left(\frac{1}{2a^2}\right)_\alpha\big[2\nabla^2\Psi_{\nu-\alpha}-\big((\nabla\Psi)^2\big)_{\nu-\alpha}\big]+\frac{3}{2}\big(e^{-2(\Phi+\Psi)}\big)_{\nu-\alpha}\big(\chi^2\big)_\alpha=\frac{1}{2\Mpl^2}\big(e^{-2\Psi}\big)_{\nu-\alpha}\Big[\rho_\alpha+\big(\mathcal{S}_\Psi\big)_\alpha\Big]\,,	
}
where we have defined $\chi\equiv H-\dot \Psi$ (which implies $\chi_\nu=H_\nu-\dot{\Psi}_\nu-i\nu m \Psi_\nu$) and 
\eqa{
&\tilde{\mathcal{D}}_\nu =\frac{m}{2}(1-e^{2\Phi})_\nu+\frac{\big(e^{2(\Phi+\Psi)}\big)_{\nu-\alpha-\beta}}{2m}\left(\frac{1}{a^2}\right)_\beta\left(\delta_{\alpha,0}\nabla^2+\nabla(\Phi_\alpha-\Psi_\alpha).\nabla\right)-\delta_{\nu,0}\frac{i}{a_s}\vec{\sigma}_s \cdot \nabla\\\nonumber&\hspace{10mm}+\frac{i}{2}(3\chi_\nu-\dot{\Phi}_\nu-i\nu m\Phi_\nu)\\
&\big(e^{imt}\mathcal{J}\big)_{\nu}=\frac{1}{24m^2}\Bigg[\lambda\Big(({\psi}^3)_{\nu+2}+({\psi}^3)^*_{4-\nu}+3{\psi}^*_{-\alpha}({\psi}^2)_{\nu-\alpha}+3{\psi}_{\alpha}({\psi}^2)^*_{2+\alpha-\nu}\Big)\\\nonumber&\hspace{32mm}+\frac{\kappa}{40m\Lambda^2}\Big(({\psi}^5)_{\nu+4}+({\psi}^5)^*_{6-\nu}+5{\psi}^*_{-\alpha}({\psi}^4)_{\nu+2-\alpha}+5{\psi}_{\alpha}({\psi}^4)^*_{4+\alpha-\nu}\\\nonumber&\hspace{82mm}+10({\psi}^2)_{\alpha}({\psi}^3)^*_{2+\alpha-\nu}+10({\psi}^2)^*_{-\alpha}({\psi}^3)_{\nu-\alpha}\Big)\Bigg]\\
&\big(\mathcal{S}_\Phi\big)_\nu=-\frac{m}{2}\Big(\big(\psi^2\big)_{\nu+2}+\big(\psi^2\big)^*_{2-\nu}+2\psi^*_{-\alpha}\psi_{\nu-\alpha}\Big)\\\nonumber&\hspace{15mm}-m\big(e^{-2(\Phi+\Psi)}\big)_{\nu-\alpha}\Big(\big(\psi^2\big)_{\alpha+2}+\big(\psi^2\big)^*_{2-\alpha}-2\psi^*_{-\beta}\psi_{\alpha-\beta}\Big)\\\nonumber
&\hspace{15mm}-\frac{\lambda}{48m^2}\Big[({\psi}^4)_{\nu+4}+({\psi}^4)^*_{4-\nu}+4{\psi}^*_{-\alpha}({\psi}^3)_{\nu+3-\alpha}+4{\psi}_{\alpha}({\psi}^3)^*_{3+\alpha-\nu}+6({\psi}^2)_{\alpha}({\psi}^2)^*_{\alpha-\nu}\Big]\,,
}
and
\eqa{
&\big(\mathcal{S}_\Psi\big)_\nu=\frac{m}{4}\Big(\big(\psi^2\big)_{\nu+2}+\big(\psi^2\big)^*_{2-\nu}+2\psi^*_{-\alpha}\psi_{\nu-\alpha}\Big)\\\nonumber&\hspace{15mm}-\frac{m}{4}\big(e^{-2\Phi}\big)_{\nu-\alpha}\Big(\big(\psi^2\big)_{\alpha+2}+\big(\psi^2\big)^*_{2-\alpha}-2\psi^*_{-\beta}\psi_{\alpha-\beta}\Big)\\\nonumber
&\hspace{15mm}+\frac{\lambda}{96m^2}\Big[({\psi}^4)_{\nu+4}+({\psi}^4)^*_{4-\nu}+4{\psi}^*_{-\alpha}({\psi}^3)_{\nu+3-\alpha}+4{\psi}_{\alpha}({\psi}^3)^*_{3+\alpha-\nu}+6({\psi}^2)_{\alpha}({\psi}^2)^*_{\alpha-\nu}\Big]\\\nonumber
&\hspace{15mm}+\frac{1}{4m}\left(\frac{1}{a^2}\right)_\beta\big(e^{-2\Phi}\big)_{\nu-\alpha-\beta}\Big(\big((\nabla\psi)^2\big)_{\alpha+2}+\big((\nabla\psi)^2\big)^*_{2-\alpha}+2\nabla\psi^*_{-\gamma}\nabla\psi_{\alpha-\gamma}\Big)\,.
}
Note that the mode decomposition of exponential functions must be understood as the mode decomposition of its Taylor expansion. For example, we have $\big(e^{2\Phi}\big)_\nu=\delta_{\nu,0}+2\Phi_\nu+2\big(\Phi^2\big)_\nu+\dots$.

The other equation we need for a closed system is the one for the vector mode. However, at the working order, vector modes are treated only at leading order and no correction is needed for them. As a result, the effective equation for the vector slow mode is simply Eq.~\eqref{eqsigma} with all variables replaced by their slow modes.

As the next step, we need to solve for the nonzero modes perturbatively. For inhomogeneous variables, however, the relevant equations are not necessarily algebraic and can be differential equations including spatial derivatives. Whereas one can write a formal solution for a differential equation, for example by using inverse Laplacian operators, the solution might appear to be nonlocal. As was the case for the background equations, the nonzero modes appear as the corrections to the SP system. As a result, ``integrating out" nonzero modes might in general result in a set of equations which are nonlocal in space. The appearance of nonlocal terms is a general feature of gauge theories (including general relativity), and is not necessarily a sign of the breakdown of any fundamental physical principle. Therefore, we expect such nonlocal contributions to show up at some level in a perturbative expansion. Fortunately, up to leading-order corrections to the SP system, these nonlocal solutions do not appear in our EFT, though they do at higher orders; see Appendix \ref{app:NREFT2}.

To leading order, the resulting solutions for nonzero modes that will be needed are
\eqa{
	\psi^{(1)}_\nu&=\left(-\frac{3iH_s}{4m}-\frac{\lambda|\psi_s|^2}{16m^3}-\frac{1}{2}\Phi_s+\frac{\nabla^2}{4m^2a_s^2}\right)\psi_s^*\delta_{\nu,2}+\frac{\lambda\psi_s^3}{48m^3}\delta_{\nu,-2}-\frac{\lambda\psi_s^*{}^3}{96m^3}\delta_{\nu,4}\,,\\
	\Psi^{(1)}_\nu&=\frac{\psi_s^*{}^2-\bar{\psi}_s^*{}^2}{16m\Mpl^2}\delta_{\nu,2}+\frac{\psi_s^2-\bar{\psi}_s^2}{16m\Mpl^2}\delta_{\nu,-2}\,.
}
Next we must substitute these expressions into to the equations for the slow modes. From the Schr\"{o}dinger equation we obtain
\eq{
	\spl{
		&i\dot{\psi}_s+\tilde{\mathcal{D}}_s\psi_s-\big(e^{imt}\mathcal{J}\big)_{s}\\
		&+\tilde{\mathcal{D}}^*_{s}\psi^*_{2}+\tilde{\mathcal{D}}^*_{2}\psi^*_{s}-2\Phi_s\big(e^{imt}\mathcal{J}\big)_{s}+\dots=0\,,
	}
}
where for the operators we have
\eq{
	\spl{
		\tilde{\mathcal{D}}_s=&-m\Phi_s+\frac{3i}{2}H_s-\frac{\nabla^2}{2ma_s^2}\\
		&-m\Phi_s^2+\frac{2\Phi_s}{ma_s^2}\nabla^2-\frac{i}{2}(\dot{\Phi}_s+3\dot{\Psi}_s)-\frac{i}{a_s} \vec{\sigma}_s \cdot \nabla+\dots\,,
}}
and
\eq{
	\spl{
		\tilde{\mathcal{D}}_2&=-m\Phi_2+\frac{i}{2}(3H_2-2i\Phi_2-6im\Psi_2)+\dots\\
		&=\frac{3i}{2}H_2+3m\Psi_2+\dots\,.
}}
Note that in the expression for $\tilde{\mathcal{D}}_s$, terms that appear on the second line are one order smaller than those on the first line. Substituting the solutions for the nonzero modes to the above expression yields 
\eq{
	\begin{split}
		&i\dot{\psi}_s+\frac{3i}{2}H_s\psi_s+\frac{1}{2ma_s^2}\nabla^2\psi_s-m\Phi_s\psi_s-\frac{\lambda}{8m^2}|\psi_s|^2\psi_s\\
		&+\left(\frac{3\rho_s}{8m\Mpl^2}+\frac{|\bar{\psi}_s|^2}{2\Mpl^2}+\frac{|\psi_s|^2}{16\Mpl^2}-\frac{m}{2}\Phi_s^2\right)\psi_s-\frac{i}{2}\left(\dot{\Phi}_s+3\dot{\Psi}_s\right)\psi_s+\frac{\nabla^4\psi_s}{8m^3a_s^4}\\
		&+(\Phi_s+2\Psi_s)\frac{\nabla^2\psi_s}{2ma_s^2}-\frac{\nabla\Psi_s\nabla\psi_s}{2ma_s^2}-i\frac{\vec{\sigma}_s \cdot \nabla\psi_s}{a_s}+\left(\frac{17\lambda^2}{8m^2}-\frac{\kappa}{\Lambda^2}\right)\frac{|\psi_s|^4\psi_s}{96m^3}\\
		&-\frac{\lambda}{16m^4a_s^2}\left(2|\nabla\psi_s|^2\psi_s+\psi_s^2\nabla^2\psi_s^*+2|\psi_s|^2\nabla^2\psi_s+\psi_s^*(\nabla\psi_s)^2\right)+\dots=0\,.
\end{split}}
Likewise, the effective equations for the gravitational potential take the following form:
\eq{
	\label{eftPhi_full}
	\begin{split}
		&\frac{\nabla^2\Phi_s}{a_s^2}-\frac{m}{2\Mpl^2}(|\psi_s|^2-|\bar{\psi}_s|^2)\\
		&+\frac{\nabla\Phi_s\nabla(\Phi_s-\Psi_s)}{a_s^2}+3(H_s\dot{\Phi}_s+2H_s\dot{\Psi}_s+\ddot{\Psi}_s)-\frac{(\rho_s+3p_s+m|\bar{\psi}_s|^2)\Phi_s}{\Mpl^2}-\frac{m|\bar{\psi}_s|^2\Psi_s}{\Mpl^2}\\
		&+(\Phi_s+2\Psi_s)\frac{m|\psi_s|^2}{2\Mpl^2}+\frac{3}{8m\Mpl^2a_s^2}(\psi_s\nabla^2\psi_s^*+\psi_s^*\nabla^2\psi_s)-\frac{\lambda}{8m^2\Mpl^2}(|\psi_s|^4-|\bar{\psi}_s|^4)+\dots=0\,,
\end{split}}
and 
\eq{
	\label{eftPsi_full}
	\begin{split}
		&\frac{\nabla^2\Psi_s}{a_s^2}-\frac{m}{2\Mpl^2}(|\psi_s|^2-|\bar{\psi}_s|^2)\\
		&-\frac{(\nabla\Psi_s)^2}{2a_s^2}-3H_s\dot{\Psi}_s-\frac{(\rho_s+m|\bar{\psi}_s|^2)\Phi_s}{\Mpl^2}-\frac{m|\bar{\psi}_s|^2\Psi_s}{\Mpl^2}+(\Phi_s+2\Psi_s)\frac{m|\psi_s|^2}{2\Mpl^2}-\frac{|\nabla\psi_s|^2}{4m\Mpl^2a_s^2}\\
		&-\frac{\lambda}{32m^2\Mpl^2}(|\psi_s|^4-|\bar{\psi}_s|^4)+\dots=0\,.
\end{split}}

All the equations obtained here will be reduced to our main results in Sec.~\ref{eftnr} if one uses $|\Phi-\Psi| \sim \order{\epsilon^2}$ to replace $\Psi$ with $\Phi$ in several places, as long as the difference does not contribute to the EFT at the current working order. However, keeping the two gravitational potentials distinct helps us to remove one parameter from the theory in the case of spherically symmetric solitons. In this case, as discussed in Sec.~\ref{sec:spherical}, we use the ansatz $\psi =f(r)e^{i\mu t}$. After neglecting the expansion of the universe, we take the following steps to remove $\mu$ from all equations: We first remove higher-order spatial derivatives that appear in the EFT equations by using lower-order equations. The result will contain higher-order time derivatives instead. However, dealing with time derivatives is not a problem as a result of the simple ansatz we consider. Then, we use the following set of redefinitions to completely remove $\mu$ from the effective equations: 
\ba 
\label{redef}
 f(r)=\left(1-\frac{\mu}{2m}\right)\tilde{f}(r)\, , \qquad 	\Phi_s=\tilde{\Phi}_s-\frac{\mu}{m}-\frac{\mu^2}{2m^2}-\frac{\mu^3}{3m^3}.
\ea
After this redefinition we have $\tilde \Phi-\Psi \sim \mu/m\sim \order{\epsilon}$, which justifies our attempt to keep the two gravitational potentials distinct even at this order. In Appendix \ref{app:NREFT2} we present the results after this redefinition, where we also go one order higher in our perturbative expansion.

\section{Effective equations with $\mathcal O(\epsilon^2)$ corrections} 
\label{app:NREFT2}
In this Appendix we provide the time-independent effective equations including $\mathcal O(\epsilon^2)$ corrections for a spherically symmetric spacetime, which can be used to further justify the effective field theory through the mass-radius relation. As stated in Sec.~\ref{sec:spherical}, here we ignore the vector and tensor components of metric perturbations as well as the expansion of the universe. Also for simplicity, we set $\kappa=0$. To achieve these results, we again remove $\mu$ from all equations using the procedure outlined at the end of Appendix \ref{app:EFT_details}, including the redefinitions of Eq.~\eqref{redef}. The effective equations that are shown here are the ones that we use for our numerical solutions. Since we have already discussed the steps required to obtain the EFT, we only present the final results here.

The effective equation for the profile of the field $\tilde{f} (r)$ is
\eq{
\label{sch2}
\spl{
&\frac{\nabla^2\tilde{f}}{2m}-m\left(1-\textcolor{black}{\tilde{\Phi}_s}-\textcolor{black}{2\Psi_s}+\textcolor{black}{2\Psi_s^2}+\textcolor{black}{2\Psi_s\tilde{\Phi}_s}+\textcolor{black}{\frac{2}{3}\tilde{\Phi}_s^2}\right)\tilde{\Phi}_s\tilde{f}+\textcolor{black}{3m\Psi_2^{(2)}\tilde{f}}+\left(\textcolor{black}{1}-\textcolor{black}{2\Psi_s}-\textcolor{black}{\frac{4}{3}\tilde{\Phi}_s}\right)\textcolor{black}{\frac{3\tilde{f}^3}{16\Mpl^2}}\\
&+\textcolor{black}{\frac{1}{2m}\nabla\tilde{f} \cdot \nabla(\tilde{\Phi}_s-\Psi_s+\Phi_2^{(1)})}-\textcolor{black}{\frac{\tilde{f}(\nabla\tilde{f})^2}{16m^2\Mpl^2}}-\left(1-\textcolor{black}{2\Psi_s}+\textcolor{black}{2\Psi_s^2}+\textcolor{black}{\Phi_2^{(1)}}-\textcolor{black}{\frac{\tilde{f}^2}{8m\Mpl^2}}\right)\frac{\lambda\tilde{f}^3}{8m^2}\\
&-\left(\textcolor{black}{1}-\textcolor{black}{2\Psi_s}+\textcolor{black}{\frac{3}{2}\tilde{\Phi}_s}\right)\textcolor{black}{\frac{\lambda^2\tilde{f}^5}{768m^5}}+\textcolor{black}{\frac{\lambda^2\tilde{f}^3(\nabla\tilde{f})^2}{1024m^7}}+\textcolor{black}{\frac{\lambda^3\tilde{f}^7}{73728m^8}}+\dots=0\,,
}}
and the equations for the gravitational potentials are given by
\eq{
\label{psi2}
	\spl{
		&\nabla^2\Psi_s-\textcolor{black}{\frac{1}{2}(\nabla\Psi_s)^2}-\left(1-\textcolor{black}{\tilde{\Phi}_s}-\textcolor{black}{2\Psi_s}+\textcolor{black}{2\Psi_s^2}+\textcolor{black}{2\tilde{\Phi}_s\Psi_s}+\textcolor{black}{\tilde{\Phi}_s^2}+\textcolor{black}{\Phi_2^{(1)}}-\textcolor{black}{\frac{3\tilde{f}^2}{32m\Mpl^2}}\right)\frac{m\tilde{f}^2}{2\Mpl^2}\\
		&-\textcolor{black}{\frac{(\nabla\tilde{f})^2}{4m\Mpl^2}}-(\textcolor{black}{1}-\textcolor{black}{2\Psi_s})\textcolor{black}{\frac{\lambda\tilde{f}^4}{32m^2\Mpl^2}}-\textcolor{black}{\frac{13\lambda^2\tilde{f}^6}{18432m^3\Mpl^4}}+\dots=0\,,
	}
} 
and
\eq{
\label{phi2}
	\begin{split}
		&\nabla^2\tilde{\Phi}_s-\left(1-\textcolor{black}{4\tilde{\Phi}_s}-\textcolor{black}{2\Psi_s}+\textcolor{black}{2\Psi_s^2}+\textcolor{black}{8\tilde{\Phi}_s\Psi_s}+\textcolor{black}{4\tilde{\Phi}_s^2}+\textcolor{black}{\Phi_2^{(1)}}+\textcolor{black}{\frac{3\tilde{f}^2}{16m\Mpl^2}}\right)\frac{m\tilde{f}^2}{2\Mpl^2}\\
		&+\textcolor{black}{\nabla\tilde{\Phi}_s \cdot \nabla(\tilde{\Phi}_s-\Psi_s)}+(\textcolor{black}{1}-\textcolor{black}{2\Psi_s})\textcolor{black}{\frac{\lambda\tilde{f}^4}{16m^2\Mpl^2}}-\textcolor{black}{\frac{\lambda^2\tilde{f}^6}{18432m^3\Mpl^4}}+\dots=0\, .
\end{split}} 
The two nonzero modes, $\Psi_{2}^{(2)}$ and $\Phi_2^{(1)}$, can be expressed in terms of the slow modes via
\eq{
\label{psi22}
\p_r\Psi_2^{(2)}=\frac{\tilde{f}^2}{16m\Mpl^2}\p_r\tilde{\Phi}_s\,,
}
and
\eq{
\label{phi21}
\nabla^2\Phi_{2}^{(1)}=12m^2\Psi_2^{(2)}-\frac{m\tilde{f}^2}{2\Mpl^2}\tilde{\Phi}_s-\frac{\lambda\tilde{f}^4}{64m^4\Mpl^2}\,
}
which shows that the solutions would be nonlocal; hence we do not attempt to substitute them into the effective equations. Finally, the mass formula of Eq.~\eqref{mass} becomes
\eq{
\spl{
M=\int_{0}^{\infty}\dd{r}4\pi r^2\Bigg[&m\tilde{f}^2\bigg(1-\textcolor{black}{\tilde{\Phi}_s}-\textcolor{black}{\frac{5}{2}\Psi_s}+\textcolor{black}{\frac{25}{8}\Psi_s^2}+\textcolor{black}{\frac{5}{2}\tilde{\Phi}_s\Psi_s}+\textcolor{black}{\tilde{\Phi}_s^2}+\textcolor{black}{\Phi_2^{(1)}}-\textcolor{black}{\frac{3\tilde{f}^2}{32m\Mpl^2}}\bigg)\\
&+\textcolor{black}{\frac{(\nabla\tilde{f})^2}{2m}}\left(\textcolor{black}{1}-\textcolor{black}{\frac{1}{2}\Psi_s}\right)+\textcolor{black}{\frac{\lambda\tilde{f}^4}{16m^2}}\left(\textcolor{black}{1}-\textcolor{black}{\frac{5}{2}\Psi_s}\right)+\textcolor{black}{\frac{13\lambda^2\tilde{f}^6}{9216m^3\Mpl^2}} \Bigg]\,.
}} 

\section{Coordinate transformation}\label{app:coordinate}
In this Appendix we discuss the coordinate transformation between the two most common metric tensors that describe a spherically symmetric spacetime, the isotropic coordinates (given by Eq.~\eqref{metric1}, neglecting vector and tensor modes) and the spherical coordinates
\eq{
	\label{spherical_coordinate}
	\dd{s}^2=-e^{2\alpha(\eta,\rho)}\dd{\eta}^2+e^{-2\beta(\eta,\rho)}\dd{\rho}^2+\rho^2\dd{\Omega}^2\,.
} 
For any physical solutions we must demand $\beta(\eta,0)=0$ to avoid singularities at the origin, which makes the spherical coordinates especially useful for numerical shooting procedures to solve the KGE equations \cite{Alcubierre:2003sx}.

Since our effective field theory in this paper is formalized in isotropic coordinates, it is useful to present the transformation from spherical coordinates to isotropic ones. To do this, we first identify the ansatzes
\eq{
	\label{rrho}
	\rho(t,r)=r\,e^{-\Psi(t,r)}\,,\quad
	\eta(t,r)=t+\Theta(t,r)\,.
}
By substituting these into Eq.~\eqref{spherical_coordinate} we obtain constraints for three unknown functions, $\Phi$, $\Psi$ and $\Theta$,
\eqa{
	\label{constraint_00}
	g_{00} &= -e^{2\Phi} = -e^{2\alpha}(1+\dot{\Theta})^2 + e^{-2\beta-2\Psi}r^2\dot{\Psi}^2\,,\\
	\label{constraint_10}
	g_{10} &= 0 = -e^{2\alpha}(1+\dot{\Theta})\Theta' - e^{-2\beta-2\Psi} (1-r\Psi') \dot{\Psi} \,,\\
	\label{constraint_11}
	g_{11} &= e^{-2\Psi} =  -e^{2\alpha}\Theta'{}^2+e^{-2\beta-2\Psi}  (1-r\Psi')^2 ~,
}
where dots and primes denote differentiation with respect to $t$ and $r$ respectively. If we expand the gravitational potentials in terms of the cosine series, for example,
\begin{align}
\Phi(t,r) = \frac{1}{2}\Phi_0(r) + \Phi_2(r)\cos(2\omega t) + \Phi_4(r)\cos(4\omega t) + \cdots ~, 
\end{align}
and the time difference in terms of the sine series,
\begin{align}
\Theta(t,r) = \Theta_2(r)\sin(2\omega t) + \Theta_4(r)\sin(4\omega t) + \cdots ~,
\end{align}
then we can solve each mode order by order. Note that $\Phi_0\sim\Psi_0 \gg \Phi_2$ while $\alpha_0\sim \beta_0 \sim \alpha_2$ \cite{Zhang:2020ntm}. Here we provide the leading-order coordinate transformations,
\begin{align}
\Phi_0(r) &= \alpha_0(r) + \mathcal O (\epsilon^2) ~,\\
\Theta_2(r)& = -\frac{1}{2\omega}\alpha_2(r) + \mathcal O(\epsilon^2) ~,\\
\Phi_0'(r) &= -\frac{\beta_0(r)}{r} + \mathcal O(\epsilon^2) ~,
\end{align}
where the first two equations are given by the coefficients of the $\cos(0\omega t)$ and $\cos(2\omega t)$ terms in Eq.~\eqref{constraint_00} and the third equation is given by the coefficient of the $\cos(0\omega t)$ term in Eq.~\eqref{constraint_11}. The derivation of higher-order transformations is straightforward.

\bibliography{references} 
\bibliographystyle{JHEP}

\end{document}